\documentclass[english,aps,prb,superscriptaddress,showpacs,showkeys,twocolumn]{revtex4-1}
\usepackage[T1]{fontenc}
\usepackage{xcolor}
\usepackage{pdfcolmk}
\usepackage{babel}
\usepackage{array}
\usepackage{booktabs}
\usepackage{units}
\usepackage{multirow}
\usepackage{amstext}
\usepackage{graphicx}
\PassOptionsToPackage{normalem}{ulem}
\usepackage{ulem}
\usepackage[unicode=true,
 bookmarks=true,bookmarksnumbered=false,bookmarksopen=false,
 breaklinks=false,pdfborder={0 0 1},backref=false,colorlinks=true]
 {hyperref}
\hypersetup{pdftitle={Effect of lattice distortion on U magnetic moments in U4Ru7Ge6 studied by polarized neutron diffraction},
 pdfauthor={M. Valiska, M. Klicpera, P. Dolezal, O. Fabelo, A. Stunault, M. Divis, V. Sechovsky},
 pdfkeywords={Itinerant 5f-electron ferromagnetism, polarized neutron diffraction, distortion, spin-orbit coupling}}

\makeatletter

\providecommand{\tabularnewline}{\\}
\providecolor{lyxadded}{rgb}{0,0,1}
\providecolor{lyxdeleted}{rgb}{1,0,0}

\DeclareRobustCommand{\lyxsout}[1]{\ifx\\#1\else\sout{#1}\fi}

\makeatother

\begin{document}

\title{Effect of lattice distortion on U magnetic moments in $\mathrm{U_{4}Ru_{7}Ge_{6}}$
studied by polarized neutron diffraction}

\author{Michal Vali\v{s}ka}
\email{michal.valiska@gmail.com}

\affiliation{Faculty of Mathematics and Physics, Charles University, DCMP, Ke
Karlovu 5, CZ-12116 Praha 2, Czech Republic}

\affiliation{Institut Laue Langevin, 71 Avenue des Martyrs, CS 20156, F-38042
Grenoble Cedex 9, France}

\author{Milan Klicpera}

\affiliation{Faculty of Mathematics and Physics, Charles University, DCMP, Ke
Karlovu 5, CZ-12116 Praha 2, Czech Republic}

\author{Petr Dole\v{z}al}

\affiliation{Faculty of Mathematics and Physics, Charles University, DCMP, Ke
Karlovu 5, CZ-12116 Praha 2, Czech Republic}

\author{Oscar Fabelo}

\affiliation{Institut Laue Langevin, 71 Avenue des Martyrs, CS 20156, F-38042
Grenoble Cedex 9, France}

\author{Anne Stunault}

\affiliation{Institut Laue Langevin, 71 Avenue des Martyrs, CS 20156, F-38042
Grenoble Cedex 9, France}

\author{Martin Divi\v{s}}

\affiliation{Faculty of Mathematics and Physics, Charles University, DCMP, Ke
Karlovu 5, CZ-12116 Praha 2, Czech Republic}

\author{Vladimír Sechovský}

\affiliation{Faculty of Mathematics and Physics, Charles University, DCMP, Ke
Karlovu 5, CZ-12116 Praha 2, Czech Republic}
\begin{abstract}
In a cubic ferromagnet, small spontaneous lattice distortions are
expected below the Curie temperature, but the phenomenon is usually
neglected. This study focuses on such an effect in the $\mathrm{U_{4}Ru_{7}Ge_{6}}$
compound. Based on DFT calculations, we propose a lattice distortion
from the cubic $Im-3m$ space group to a lower, rhombohedral, symmetry
described by the $R-3m$ space group. The strong spin-orbit coupling
of the uranium ions plays an essential role in lowering the symmetry,
giving rise to two different U sites (U1 and U2). Using polarized
neutron diffraction in applied magnetic fields of $\unit[1]{T}$ and
$\unit[9]{T}$ in the ordered state ($\unit[1.9]{K}$) and in the
paramagnetic state ($\unit[20]{K}$), we bring convincing experimental
evidence of this splitting of the U sites, with different magnetic
moments. The data have been analyzed both by maximum entropy calculations
and by a direct fit in the dipolar approximation. In the ordered phase,
the $\mu_{\mathrm{L}}/\mu_{\mathrm{S}}$ ratio of the orbital and
spin moments on the U2 site is remarkably lower than for the free
$\mathrm{U^{3+}}$ or $\mathrm{U^{4+}}$ ion , which points to a strong
hybridization of the U $5f$ wavefunctions with the $4d$ wavefunctions
of the surrounding Ru. On the U1 site, the $\mu_{\mathrm{L}}/\mu_{\mathrm{S}}$
ratio exhibits an unexpectedly low value: the orbital moment is almost
quenched, like in metallic $\alpha$-Uranium. As a further evidence
of the $5f-4d$ hybridization in the $\mathrm{U_{4}Ru_{7}Ge_{6}}$
system, we observe the absence of a magnetic moment on the Ru1 site,
but a rather large induced moment on the Ru2 site, which is in closer
coordination with both U positions. Very similar results are obtained
at $\unit[20]{K}$ in the ferromagnetic regime induced by the magnetic
field of $\unit[9]{T}$. This shows that applying a strong magnetic
field above the Curie temperature also leads to the splitting of the
uranium sites, which further demonstrates the intimate coupling of
the magnetic ordering and structural distortion. We propose that the
difference between the magnetic moment on the U1 and U2 sites results
from the strong spin-orbit interaction with different local point
symmetries.
\end{abstract}


\maketitle

\section{Introduction}

In condensed matter, the uranium 5$f$ wavefunctions have a large
spatial extent and interact strongly with the outer electrons of the
neighboring ions leading to the loss of their atomic character and
the delocalization of the 5$f$ electrons. The outer-electron configuration
of the ligands of a U-ion, the coordination number and the ligand
sphere geometry hence play key roles in physics of U intermetallics.
The interplay between these physical characteristics gives a rise
to an abundance of exotic physical phenomena often connected with
complex 5$f$ electron magnetism and various exotic low-temperature
states. The loss of atomic character of the 5$f$ electron wavefunctions
and delocalization of 5$f$ electrons have a fatal impact on the formation
of the U magnetic moments which are then often found dramatically
reduced with respect to the $\mathrm{U^{3+}}$ and $\mathrm{U^{4+}}$
free-ion moment values. Unlike the 3$d$ transition metals, which
generally exhibit only a spin magnetic moment due to ``freezing\textquotedblright{}
of the orbital component in the crystal field, the relativistic effects,
namely the strong spin-orbit interaction, in the heavy atoms like
U induce a large orbital polarization\cite{Brooks1983,Brooks1985}
which is boosting the formation of a considerable orbital moment even
in the case of itinerant 5$f$ electrons. The relativistic energy
band calculations performed for U compounds by rule provide an orbital
component larger than the antiparallel spin component of a U magnetic
moment.\cite{Brooks1983,Brooks1985,Eriksson1990,Eriksson1990a,Severin1991,Norman1986,Norman1988}
These findings are in agreement with the results of polarized neutron
diffraction (PND) studies of U-compounds. The pioneering PND studies
were focused on $\mathrm{US}$\cite{Wedgewood1972}, $\mathrm{UO_{2}}$\cite{Lander1976a,Faber1976}
and $\mathrm{USb}$\cite{Lander1976b}, considered to be localized
5$f$ electron systems. The $e-e$ interactions of U ions with ligands
lead to a significant hybridization of the U 5$f$-electron states
with the non-5$f$ valence electron states (5$f$-ligand hybridization).
A strong evidence of the important role of the 5$f$-ligand hybridization
was first experimentally revealed by PND on $\mathrm{URh_{3}}$\cite{Delapalme1978}
with the observation of the enhanced elemental susceptibility of the
Rh ion, twice the value of metal Rh. The 5$f$-ligand hybridization
mechanism was theoretically explained for $\mathrm{U}T_{3}$ and $\mathrm{U}X_{3}$
($T=$ transition metal and $X=p$-element, respectively) compounds
by Koelling \textit{et al.}\cite{Koelling1985} and corroborated by
band structure calculations performed by Eriksson \textit{et al.}.\cite{Eriksson1989}

The delocalization of the 5$f$ electrons due to the large overlap
of the 5$f$ wavefunctions of neighboring U ions and the strong 5$f$-ligand
hybridization is accompanied by a reduction of the 5$f$ magnetic
moment. Despite this, the strong spin-orbit coupling induces a predominant
orbital magnetic moment antiparallel to the spin moment in the spin-polarized
5$f$ energy bands as first demonstrated for the itinerant ferromagnet
$\mathrm{UN}$.\cite{Brooks1983} This may lead to a very small total
U magnetic moment, no more than a few hundredths of $\mu_{\mathrm{B}}$,
as observed in the itinerant 5$f$ ferromagnet $\mathrm{UNi_{2}}$,\cite{Sechovsky1980}
and further confirmed by polarised neutrons\cite{Fournier1986} and
first principles electronic structure calculations.\cite{Severin1991}
Despite the itinerant character of the magnetism, $\mathrm{UNi_{2}}$
exhibits very strong magnetocrystalline anisotropy with $H_{\mathrm{a}}\gg\unit[35]{T}$
at $\unit[4.2]{K}$.\cite{Severin1991} 

The most prominent example of almost complete compensation of the
spin ($\mu_{\mathrm{S}}$) and orbital ($\mu_{\mathrm{L}}$) 5$f$
moments is $\mathrm{UFe_{2}}$ with $\mu_{\mathrm{L}}=\unit[0.23]{\mu_{\mathrm{B}}}$
and $\mu_{\mathrm{S}}=\unit[0.22]{\mu_{\mathrm{B}}}$, resulting in
a net moment at the U site of only $\unit[0.01]{\mu_{\mathrm{B}}}$.
This specific situation together with the considerably different spatial
distributions of the orbital and spin magnetizations result in an
unusual magnetic form factor exhibiting a maximum at a finite $q$.\cite{Wulff1989,Lebech1991}
The experimental observations fulfilled the theoretical prediction
of Brooks \textit{et al.}\cite{Brooks1988} based on the finding that
the 5$f$-ligand hybridization reduces disproportionately the orbital
moment, which usually dominates the total uranium moment in compounds,
and consequently the orbital moment was predicted to become comparable
to the spin moment in $\mathrm{UFe_{2}}$.

All the ferromagnetic uranium compounds studied so far by PND are
characterized by a single U site in the crystallographic unit cell,
both in the paramagnetic and the ferromagnetic states. The situation
may be different for $\mathrm{U_{4}Ru_{7}Ge_{6}}$. In the paramagnetic
phase, $\mathrm{U_{4}Ru_{7}Ge_{6}}$ crystallizes in the cubic $Im-3m$
space group with a single U site.\cite{Menovsky1988,Lloret1987,Mentink1991}
Our recent room temperature XRPD measurement confirmed this statement.\cite{Valiska2017}
Below $T_{\mathrm{C}}=\unit[10.7]{K}$, $\mathrm{U_{4}Ru_{7}Ge_{6}}$
orders ferromagnetically with the ground state easy magnetization
axis pointing along the $\left[111\right]$ crystallographic axis.
Thermal expansion measurements\cite{Valiska2017} show magnetostrictive
effects of the order of $10^{-6}$, and a subsequent rhombohedral
distortion. DFT calculations in the rhombohedral, ferromagnetic, ground
state predict the splitting of the single cubic U site into two non-equivalent
sites, with very different magnetic moments.\cite{Valiska2017}

In this paper we present a detailed PND study at low temperature on
a single crystal which provide microscopic experimental evidence of
the aforementioned predictions. We also present further DFT calculations
including the generalized gradient correction (GGA). 

\section{Experimental Details}

The $\mathrm{U_{4}Ru_{7}Ge_{6}}$ single crystal used in this study
was prepared by the Czochralski method in a tri-arc furnace. It was
cut to $\unit[2\times2\times2]{mm^{3}}$ cubic shape. Details of the
growth and characterization are published elsewhere, as well as the
room temperature structure determination by X-ray powder diffraction
(XRPD) on a powdered portion of the single crystal from the current
study.\cite{Valiska2017} No sign of spurious phases have been detected
either by XRPD nor by EDX analysis.\cite{Valiska2017} All the neutron
scattering experiments were carried out at the ILL, Grenoble.\cite{Valiska2016}
Polarized neutron diffraction was carried out using the D3 diffractometer.
We collected a set of 448 flipping ratios up to $\frac{\sin\theta}{\lambda}=\unit[0.7]{\textrm{Å}^{-1}}$
with incident beam polarization of 0.95 and wavelength $\lambda=\unit[0.85]{\mathring{A}}$.
To extract the magnetic structure factors from the polarized neutron
data, one must have an accurate knowledge of the nuclear structure
from the same sample at the same temperatures, including the extinction
corrections. For a quick check, we used the Laue instrument CYCLOPS.
A full unpolarized data collection was carried out at the D9 diffractometer.
At each temperature, we collected a set of 1190 reflections with $\unit[0.1]{\textrm{Å}^{-1}}<\frac{\sin\theta}{\lambda}<\unit[1.0]{\textrm{Å}^{-1}}$
using a wavelength of $\lambda=\unit[0.841]{\mathring{A}}$.

\section{Results}

\subsection{Structural study}

The room-temperature space group $Im-3m$ has the group $R-3m$ as
the only rhombohedral maximal subgroup.\cite{Aroyo2006} In the frame
of the $R-3m$ space group former single U site in a Wyckoff position
$8c$ is split into two different sites U1 and U2 with Wyckoff position
$3b$ and $9d$, respectively. As it is in agreement with the prediction
of two different U sites with the same multiplicity by DFT\cite{Valiska2017},
we assume $R-3m$ as a possible subgroup describing the ground state
structure of $\mathrm{U_{4}Ru_{7}Ge_{6}}$. All the following data
will be comparably refined using both the room temperature cubic space
group $Im-3m$ and proposed distorted rhombohedral $R-3m$ space group,
to rigorously study the ground state structure. The transformations
of the unit cell parameters, lattice vectors, atomic site fractional
coordinates and $h,k,l$ indices between the $Im-3m$ and $R-3m$
space group are summarized in Table \ref{tab:Structure}, where we
are using the hexagonal description of $R-3m$.

\begin{table*}
\begin{centering}
\begin{tabular}{cccc}
\toprule 
\multicolumn{2}{c}{$Im-3m$} & \multicolumn{2}{c}{$R-3m$, hexagonal axes}\tabularnewline
\midrule
\midrule 
\multirow{2}{*}{U $8c$} & \multirow{2}{*}{(0.25, 0.25, 0.25)} & U1 $3b$ & (0, 0, 0.5)\tabularnewline
\cmidrule{3-4} 
 &  & U2 $9d$ & (0.5, 0, 0.5)\tabularnewline
\midrule 
Ru1 $2a$ & (0,0,0) & Ru1 $3a$ & (0, 0, 0)\tabularnewline
\midrule 
Ru2 $12d$ & (0.25, 0, 0.5) & Ru2 $18g$ & ($\underline{x_{\mathrm{Ru2}}}$, 0, 0.5), $\underline{x_{\mathrm{Ru2}}}\sim0.25$ \tabularnewline
\midrule 
Ge $12e$ & ($x_{\mathrm{Ge}}$, 0, 0), $x_{\mathrm{Ge}}\sim0.31$  & Ge $18h$ & ($\underline{\frac{x_{\mathrm{Ge}}}{3}}$,$\underline{\frac{-x_{\mathrm{Ge}}}{3}}$,
$\underline{\frac{-2x_{\mathrm{Ge}}}{3}}$)\tabularnewline
\midrule 
\multicolumn{2}{c}{$a_{\mathrm{cub}},V_{\mathrm{cub}}$} &  & $\begin{array}{c}
\vec{a_{\mathrm{hex}}}=\vec{b_{\mathrm{cub}}}-\vec{a_{\mathrm{cub}}},\thinspace a_{\mathrm{hex}}=\sqrt{2}a_{\mathrm{cub}}\\
\vec{c_{\mathrm{hex}}}=\frac{1}{2}\left(\vec{a_{\mathrm{cub}}}+\vec{b_{\mathrm{cub}}}+\vec{c_{\mathrm{cub}}}\right),\thinspace c_{\mathrm{hex}}=\frac{\sqrt{3}}{2}a_{\mathrm{cub}}\\
V_{\mathrm{hex}}=\frac{3}{2}V_{\mathrm{cub}}
\end{array}$ \tabularnewline
\midrule 
\multicolumn{4}{c}{$\left(h,k,l\right)_{\mathrm{hex}}=\left(h,k,l\right)_{\mathrm{cub}}\left(\begin{array}{ccc}
-1 & 0 & \frac{1}{2}\\
1 & -1 & \frac{1}{2}\\
0 & 1 & \frac{1}{2}
\end{array}\right)$}\tabularnewline
\bottomrule
\end{tabular}
\par\end{centering}
\caption{\label{tab:Structure}Summary of the structural transformation between
$Im-3m$ and $R-3m$ spacegroup. Underlined coordinates in the $R-3m$
spacegroup are free parameters and are thus connected with the distortion.}

\end{table*}

\begin{figure}
\begin{centering}
\includegraphics[scale=0.85]{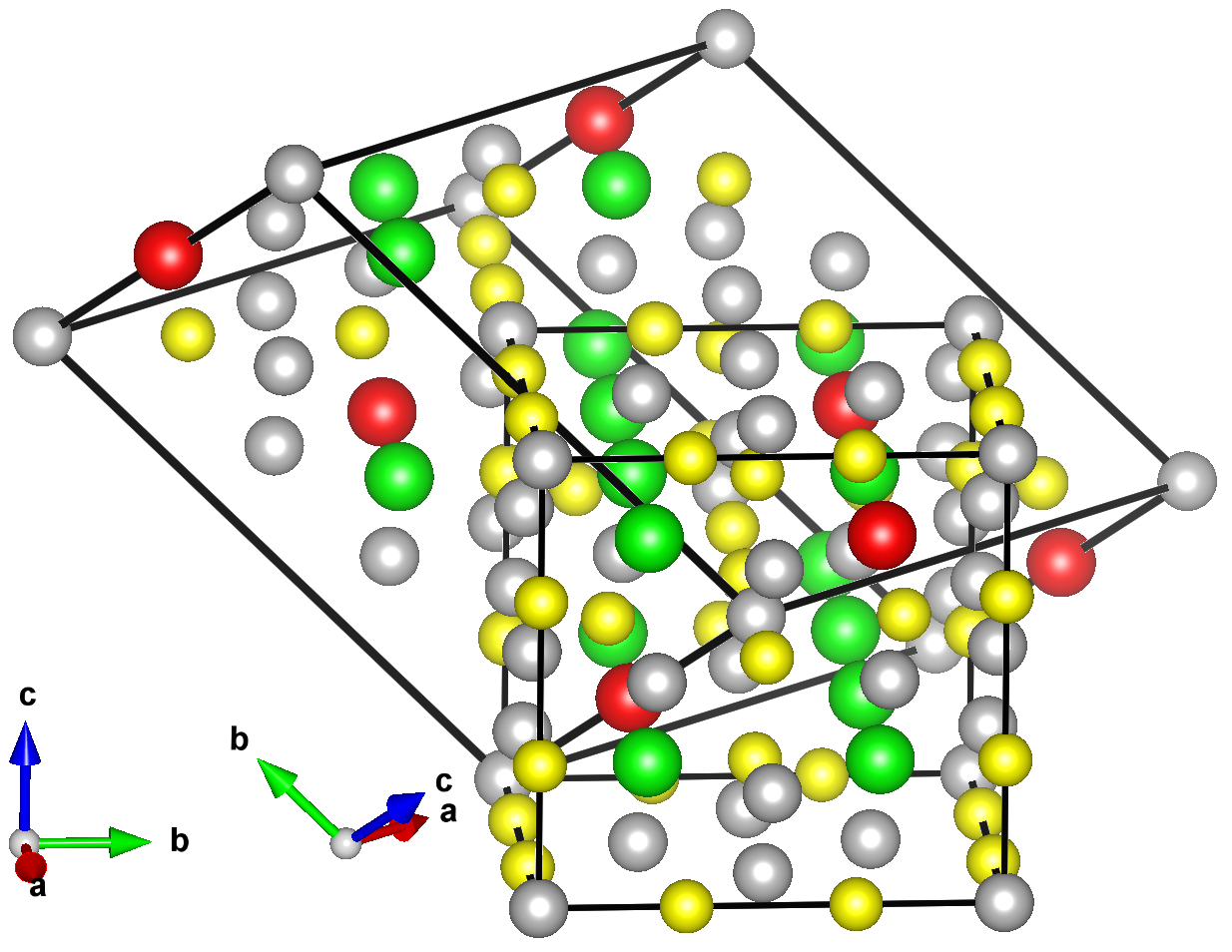}
\par\end{centering}
\caption{\label{fig:Mutual-orientation-of}Mutual orientation of the cubic
$Im-3m$ unit cell and the rhombohedral $R-3m$ in the hexagonal axes.
U1 sites are marked red, U2 green, Ru1 and Ru2 gray, and Ge are yellow. }

\end{figure}

As can be seen from the lattice vector transformation in Table \ref{tab:Structure}
the rhombohedral $R-3m$ unit cell is rotated with respect to the
cubic $Im-3m$ in the way that former $\left[111\right]_{\mathrm{cub}}$
direction is parallel with the new $\left[001\right]_{\mathrm{hex}}$
and $\left[-110\right]_{\mathrm{cub}}$ points along transformed $\left[100\right]_{\mathrm{hex}}$
(see Figure \ref{fig:Mutual-orientation-of}). The six $\mathrm{Ru2_{cubic}}$
$12d$ ions are forming a hexagonal arrangement around the U ions
in the cubic structure that is perpendicular to the body diagonals.
This local coordination is crucial for the structure of $\mathrm{U_{4}Ru_{7}Ge_{6}}$,
as the shortest inter-uranium distance $d_{\mathrm{U-U}}=\frac{a_{\mathrm{cub}}}{2}\approx\unit[4.14665]{\textrm{Å}}$
is rather large while the distance between U ion and the nearest Ru
ions $d_{\mathrm{U-Ru2_{cubic}}}\approx\unit[2.93212]{\textrm{Å}}$
highlights the importance of hybridization between U 5$f$ and Ru
4$d$ wavefunctions. The proposed symmetry change is most pronounced
in the change of the fractional coordinate $x$ of the $\mathrm{Ru2_{cubic}}$
$12d$ ($\mathrm{Ru2_{hex}}$ $18g$) site, that is no longer fixed
by symmetry. Formerly the U position $8c$ has $\mathrm{.-3m}$ point
symmetry (it has no symmetry along the cubic primary directions in
the $\left\{ 100\right\} _{\mathrm{cub}}$ family, three-fold rotoinversion
axis along the secondary direction family $\left\{ 111\right\} _{\mathrm{cub}}$
and mirror plane perpendicular to the $\left\{ 110\right\} _{\mathrm{cub}}$
family). As the fractional coordinate $x$ of the $\mathrm{Ru2_{hex}}$
position starts to deviate from the 0.25 value, this strict hexagonal
arrangement is conserved only for the U1 sites in the distorted rhombohedral
structure. It can be seen from the preserved site symmetry of the
U1 site, which is $\mathrm{-3m}$ (primary direction of the rhombohedral
lattice in the hexagonal axes is $\left\{ 001\right\} _{\mathrm{hex}}$
that is in our case parallel with the $\left\{ 111\right\} _{\mathrm{cub}}$
and secondary direction is $\left\{ 100\right\} _{\mathrm{hex}}$,
parallel with the $\left\{ 110\right\} _{\mathrm{cub}}$ ) while the
U2 site has only $\mathrm{.2/m}$ and the hexagonal arrangement of
the Ru ions is distorted.

One should notice that the proposed small distortion does not cause
the appearance of any additional Bragg reflections within the experimental
sensitivity of our experiments. It only introduces different rules
for merging of the equivalent reflections. This was confirmed using
the neutron Laue diffractometer CYCLOPS, where we did not observe
any additional reflections down to $\unit[2]{K}$. See Supplemental
Material for the Laue pictures.

The unpolarized neutron diffraction experiment gives 371(216) inequivalent
reflections assuming a $R-3m$ ($Im-3m$) space group. The internal
agreement factor of the equivalent reflections was 4.74\% (6.43\%)
at $\unit[1.9]{K}$ and 3.50\% (4.40\%) at $\unit[20]{K}$ for $R-3m$
($Im-3m$) space group. It was thus always slightly better using the
distorted $R-3m$ space group, even in the paramagnetic phase.

As the intensities at $\unit[1.9]{K}$ (ordered state) should be affected
both by nuclear and magnetic contribution we have measured the temperature
dependence of four different strong reflections from $\unit[15]{K}$
down to $\unit[1.9]{K}$ . Their intensity did not show any significant
change. Such observation is acceptable since the spontaneous magnetic
moment of $\mathrm{U_{4}Ru_{7}Ge_{6}}$ is only $\unit[0.85]{\mu_{\mathrm{B}}/f.u.}$.
We have also treated the reflections with a large enough scattering
vector, \textit{i.e.} $\frac{\sin\theta}{\lambda}>\unit[0.6]{\textrm{Å}^{-1}}$,
separately. These should be much less affected by the magnetic contribution
as was later confirmed by polarized neutron experiment where reflections
above this value showed flipping ratios close to 1 (see Section III.
C.).

The breaking of symmetry can lead to the existence of the twins (4
types of the domains in our case). We have checked this possibility
and refined our data with including these four twin components. However,
based on the twin domain fractions obtained for each twin domain $\left[\unit[97(2):0(2):0(1):3]{\%}\right]$
we can conclude that after the phase transition the sample remain
mainly as a single domain. 

\begin{table*}
\begin{centering}
\begin{tabular}{ccccccccc}
\toprule 
\multicolumn{9}{c}{$Im-3m$}\tabularnewline
\midrule
\midrule 
 & \multicolumn{5}{c}{$B_{\mathrm{iso}}\left[\textrm{Å}^{2}\right]$} & \multirow{2}{*}{$x_{\mathrm{Ge}}$} & \multirow{2}{*}{} & \multirow{2}{*}{$R_{\mathrm{F}}$}\tabularnewline
\cmidrule{1-6} 
 & \multicolumn{2}{c}{$\mathrm{U}$} & $\mathrm{Ru1}$ & $\mathrm{Ru2}$ & $\mathrm{Ge}$ &  &  & \tabularnewline
\midrule 
$\unit[20]{K}$ & \multicolumn{2}{c}{0.075(12)} & 0.120(12) & 0.130(27) & 0.120(12) & 0.3114(1) &  & 3.72\tabularnewline
\midrule 
$\unit[1.9]{K}$ & \multicolumn{2}{c}{0.066(12)} & 0.117(12) & 0.127(28) & 0.115(12) & 0.3115(1) &  & 3.66\tabularnewline
\midrule 
$\unit[1.9]{K}$, $q>\unit[0.6]{\textrm{Å}^{-1}}$ & \multicolumn{2}{c}{-0.049(10)} & 0.006(7) & 0.020(13) & 0.001(7) & 0.3115(1) &  & 3.05\tabularnewline
\midrule
\midrule 
\multicolumn{9}{c}{$R-3m$ }\tabularnewline
\midrule
\midrule 
 & \multicolumn{5}{c}{$B_{\mathrm{iso}}\left[\textrm{Å}^{2}\right]$} & \multirow{2}{*}{$x_{\mathrm{Ge}}$} & \multirow{2}{*}{$x_{\mathrm{Ru2}}$} & \multirow{2}{*}{$R_{\mathrm{F}}$}\tabularnewline
\cmidrule{1-6} 
 & $\mathrm{U}1$ & $\mathrm{U}2$ & $\mathrm{Ru1}$ & $\mathrm{Ru2}$ & $\mathrm{Ge}$ &  &  & \tabularnewline
\midrule 
$\unit[20]{K}$ & 0.060(23) & 0.106(26) & 0.090(10) & 0.094(10) & 0.039(12) & 0.3117(2) & 0.2500(1) & 4.05\tabularnewline
\midrule 
$\unit[1.9]{K}$ & 0.057(23) & 0.031(12) & 0.088(25) & 0.096(10) & 0.093(10) & 0.3114(2) & 0.2500(1) & 3.98\tabularnewline
\midrule 
$\unit[1.9]{K}$, $q>\unit[0.6]{\textrm{Å}^{-1}}$ & 0.006(17) & 0.016(11) & 0.067(18) & 0.045(10) & 0.055(10) & 0.3112(1) & 0.2498(1) & 3.32\tabularnewline
\bottomrule
\end{tabular}
\par\end{centering}
\caption{\label{tab:Structural-parameters}Structural parameters and isotropic
displacement parameters obtained from the refinement of the neutron
diffraction data. The rows labeled as $\unit[20]{K}$ and $\unit[1.9]{K}$
show results for refinement using the whole set of reflections, while
the one labeled as $\unit[1.9]{K}$, $q>\unit[0.6]{\textrm{Å}^{-1}}$shows
results only for the reflections with $q>\unit[0.6]{\textrm{Å}^{-1}}$.}
\end{table*}

The structure was refined from the measured integrated intensities,
corrected for absorption, using the Fulprof software package \cite{Roisnel2000,Rodriguez-Carvajal1993},
including extinction corrections and isotropic temperature factors.
The results are summarized in Table \ref{tab:Structural-parameters},
with different agreement factors for the cubic ($Im-3m$) and rhombohedral
($R-3m$) space groups. Negative atomic displacement parameter $B_{\mathrm{iso}}$
for U ion at $\unit[1.9]{K}$, when only reflections with $\frac{\sin\theta}{\lambda}>\unit[0.6]{\textrm{Å}^{-1}}$
are assumed, is consistent with our model of loss of the cubic symmetry
in the ground state. Nevertheless, it can not be taken as a clear
evidence for the structure change. A possible site mixing was also
checked with no significant effect on the agreement factors. The crucial
importance of the performed neutron diffraction study was the structure
determination as far as possible, including the extinction and absorption
correction, on the same crystal and at the same temperature as in
the following polarized neutron experiment.

\subsection{DFT calculations}

Details of our calculations can be found in our previous work.\cite{Valiska2017}
To obtain the microscopic information about the values of spin and
orbital magnetic moments we applied the methods based on density functional
theory (DFT). To solve Kohn-Sham-Dirac 4-components equations we used
the latest version of the computer code full potential local orbitals
(FPLO).\cite{Koepernik} We used several $k$-meshes in the Brillouin
zone to ensure the convergence of charge densities, total energy and
magnetic moments. For the sake of simplicity we assume the ferromagnetic
arrangement of the magnetic structure and the local spin density (LSDA)\cite{Perdew1992}
and generalized gradient correction (GGA)\cite{Perdew1996} were used
for exchange and correlation contribution to the total energy. The
5$f$ states of uranium were treated as itinerant Bloch states. 

To further test the performance of LSDA and GGA the equilibrium volume
were calculated and compared with experimental value $V_{0}$. The
LSDA overbinds the minimal volume by $\unit[4.8]{\%}$ ($V/V_{0}=95.2$).
In contrast to this value the GGA\cite{Perdew1996} improves the agreement
with experiment remarkably ($V/V_{0}=1.01$). Therefore the GGA description\cite{Perdew1996}
of $\mathrm{U_{4}Ru_{7}Ge_{6}}$ seems to be more precise than LSDA.\cite{Perdew1992}
Calculated components of magnetic moment using GGA for all the atoms
in the unit cell are listed in the Table \ref{tab:Magnetic-moments-calculated}.

\subsection{Polarized neutron diffraction study}

The flipping ratio method using a polarized neutron beam is a powerful
tool to study small magnetic moments. It is based on the measurement
of the flipping ratio $R=\frac{I^{+}}{I^{-}}$ of scattered intensities
$I^{+}$ and $I^{-}$, with primary beam polarized parallel (+) or
antiparallel (-) to the applied vertical field direction. The magnetic
easy axis of $\mathrm{U_{4}Ru_{7}Ge_{6}}$ at the ground state is
$\left[111\right]_{\mathrm{cub}}$ (i.e. $\left[001\right]_{\mathrm{hex}}$
).\cite{Valiska2017} Our sample was thus aligned to have the external
vertical magnetic field parallel to that direction. We collected a
set of flipping ratios at the same temperatures ($\unit[1.9]{K}$
and $\unit[20]{K}$) as in the unpolarized experiment. The applied
magnetic field was $\unit[1]{T}$ and $\unit[9]{T}$ for both measured
temperatures. 

The obtained results were treated both with respect to cubic $Im-3m$
and rhombohedral $R-3m$ space group. While merging the equivalent
flipping ratios obtained in $\unit[9]{T}$ and $\unit[1.9]{K}$ within
rhombohedral $R-3m$ space group leads to the 78 independent values
with internal agreement factor of 1\%, the same approach for cubic
$Im-3m$ space group gives 52 independent values with a degraded 3.5
\% internal agreement factor. The inadequacy of the cubic description
can be further illustrated by the example of the two reflections with
dramatically different flipping ratios. These are namely $\left(030\right)_{\mathrm{hex}}$
reflection with $R=1.28(2)$ and $\left(211\right)_{\mathrm{hex}}$
with $R=0.79(2)$ as two inequivalent ones in the $R-3m$ space group.
But according to the $Im-3m$ space group, they should be equivalent
within the $\left\{ 2-11\right\} _{\mathrm{cub}}$ family. It clearly
shows, that cubic space group $Im-3m$ cannot be used to describe
the ground state of $\mathrm{U_{4}Ru_{7}Ge_{6}}$. Same result was
observed for the data obtained at $\unit[1.9]{K}$ in $\unit[1]{T}$
and at $\unit[20]{K}$ and $\unit[9]{T}$. We will thus focus only
on the description using rhombohedral $R-3m$ space group in following
treatment of polarized neutron data, both using maximum entropy calculations
or direct flipping ratios refinement.

\subsubsection{Maximum entropy method (MAXENT)}

\begin{figure*}
\begin{centering}
\includegraphics[scale=0.14]{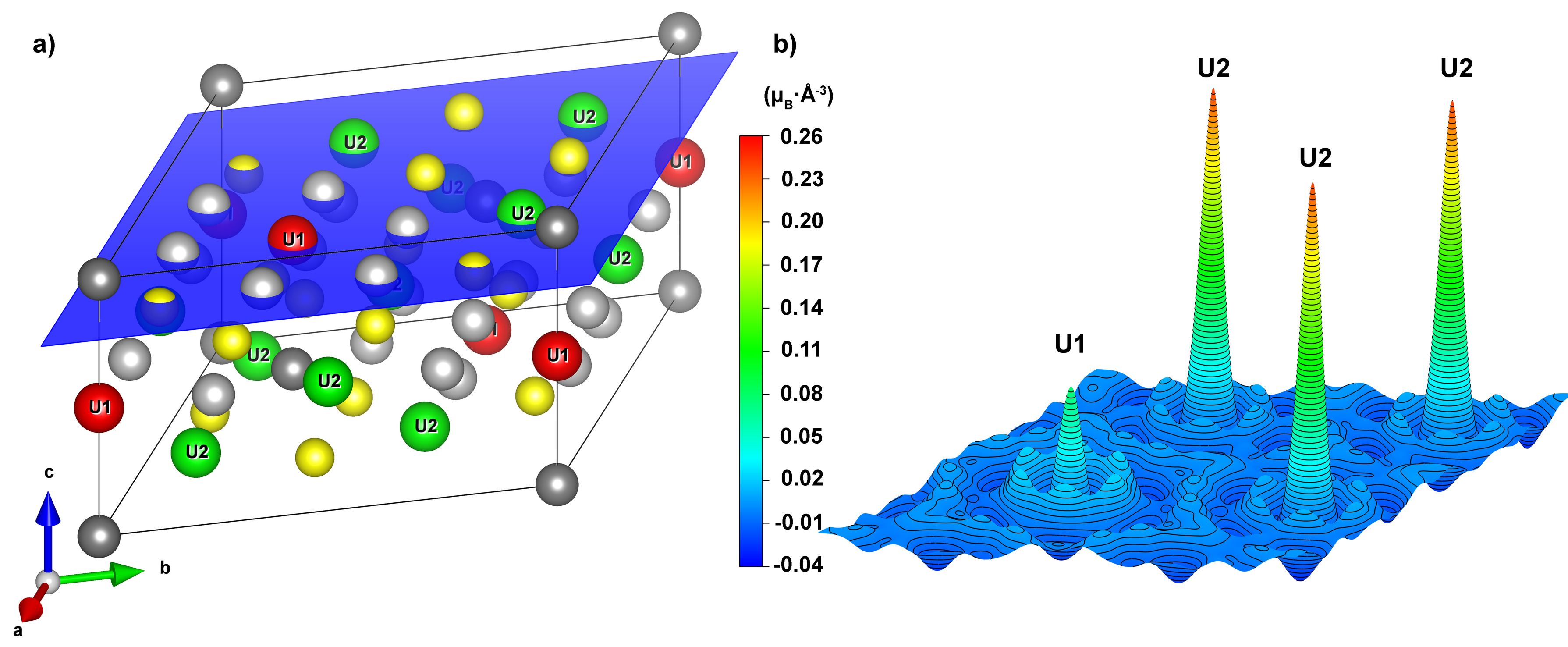}
\par\end{centering}
\caption{\label{fig:MAXENT}a) indicated section in the $R-3m$ space group
representation, that is perpendicular to the $\left[001\right]_{\mathrm{hex}}$
axis, b) corresponding MAXENT magnetization density map measured in
$\unit[9]{T}$ and $\unit[1.9]{K}$. Contour lines are at $\unit[0.004]{\mu_{\mathrm{B}}}\cdot\textrm{Å}^{-3}$
.}

\end{figure*}

The maximum entropy approach is not affected by any prior assumption
of the distribution of magnetic density. Its only inputs are the symmetry
and dimensions of the unit cell and the magnetic structure factors
obtained from the measured flipping ratios. In our case the whole
unit cell was divided in to 235 x 235 x 145 separated voxels. We used
Dysnomia software utilizing the Cambridge algorithm \cite{Momma2013}
to calculate the most probable spin density map. The initial state
was a flat magnetic density distribution over the unit cell. The results
are plotted using the VESTA software. \cite{Momma2011} The resulting
three-dimensional spin density map agrees with the experimental magnetic
structure factors and has maximal entropy. Figure \ref{fig:MAXENT}
shows the spin density map obtained at $\unit[9]{T}$ and $\unit[1.9]{K}$
in a slice perpendicular to the $\left[001\right]_{\mathrm{hex}}$
axis at the fractional coordinate $z\approx0.833333$. This slice
truncates both the U1 and U2 ions and evidences a density almost three
times higher on the U2 sites than on the U1 site: $\unit[\sim0.24]{\mu_{\mathrm{B}}\cdot\textrm{Å}^{-3}}$
and $\unit[\sim0.078]{\mu_{\mathrm{B}}\cdot\textrm{Å}^{-3}}$ respectively,
at the center. Integration in the spherical region around the given
atomic position can serve as a rough estimation of the magnetic moments
associated to each site. We have performed integration in the sphere
with gradually increasing radius. The obtained magnetic moments showed
saturation around $\unit[\sim1.8]{\textrm{Å}}$ for both the U1 and
U2 positions. This value is close to the experimental atomic radius
of $\unit[1.75]{\textrm{Å}}$ of uranium.\cite{Slater1964} See Supplemental
Material for the integrated magnetic moment as a function of the radius.
Integrated values are $\unit[0.17(1)]{\mu_{\mathrm{B}}}$ and $\unit[0.22(1)]{\mu_{\mathrm{B}}}$
for the U1 and U2 site at $\unit[9]{T}$ and $\unit[1.9]{K}$, respectively.
Qualitatively comparable results were obtained for the measurement
at $\unit[1.9]{K}$ and $\unit[1]{T}$ and at $\unit[20]{K}$, and
$\unit[9]{T}$. See Supplemental Material for the corresponding magnetization
density maps. All values are summarized in Table \ref{tab:Magnetic-moments-calculated}.

\subsubsection{Direct refinement of the flipping ratios}

To describe the magnetic structure of $\mathrm{U_{4}Ru_{7}Ge_{6}}$
in more detail and distinguish between the spin $\mu_{\mathrm{S}}$
and orbital $\mu_{\mathrm{L}}$ components of the magnetic moments
we compared the measured flipping ratios (or magnetic structure factors)
with a model, in the dipolar approximation. Our least squares refinement
was performed using the FullProf/WinPlotr\cite{Roisnel2000,Rodriguez-Carvajal1993}
software. The dipolar approximation uses an isotropic description
of the magnetic form factors $f_{\mathrm{M}}\left(q\right)$ and can
be written:

\begin{equation}
\begin{array}{c}
f_{\mathrm{M}}\left(q\right)=\mu\left(\left\langle j_{0}\left(q\right)\right\rangle +C_{2}\left\langle j_{2}\left(q\right)\right\rangle \right)\\
\\
C_{2}=\frac{\mu_{\mathrm{L}}}{\mu}=\frac{\mu_{\mathrm{L}}}{\mu_{\mathrm{L}}+\mu_{\mathrm{S}}}
\end{array}
\end{equation}

Using tabulated values of the $\left\langle j_{0}\left(q\right)\right\rangle $
and $\left\langle j_{2}\left(q\right)\right\rangle $ radial integrals,\cite{Brown}
we could extract the spin $\mu_{\mathrm{S}}$ and orbital $\mu_{\mathrm{L}}$
magnetic moments. We used the $\mathrm{Ru^{1+}}$ values for ruthenium
and either the $\mathrm{U^{3+}}$ or the $\mathrm{U^{4+}}$ values
for uranium (although the actual valence of uranium in $\mathrm{U_{4}Ru_{7}Ge_{6}}$
may be different). 

The assumed ground-state space group $R-3m$ of $\mathrm{U_{4}Ru_{7}Ge_{6}}$
has two different U and Ru sites but, according to the symmetry, every
$f_{\mathrm{M}}\left(q_{h,k,l}\right)$ reflection has a contribution
from all four sites, and one has to include them all at once into
the least squares fit. Best fits for the all experimental conditions
were obtained assuming a $\mathrm{U^{3+}}$ form factor, although
the difference between $\mathrm{U^{3+}}$ and $\mathrm{U^{4+}}$ form
factor is very small.\cite{Freeman1976} The results are summarized
in Table \ref{tab:Magnetic-moments-calculated}. As an example, the
measured and calculated flipping ratios are plotted in Figure \ref{fig:Flipping-ratios-measured}
for the ($\unit[9]{T}$, $\unit[1.9]{K}$) data set. See Supplemental
Material for the comparison of the measured and calculated flipping
ratios for measurement at $\unit[1.9]{K}$ and $\unit[1]{T}$ and
at $\unit[20]{K}$ and $\unit[9]{T}$. 

\begin{figure}
\begin{centering}
\includegraphics[scale=0.36]{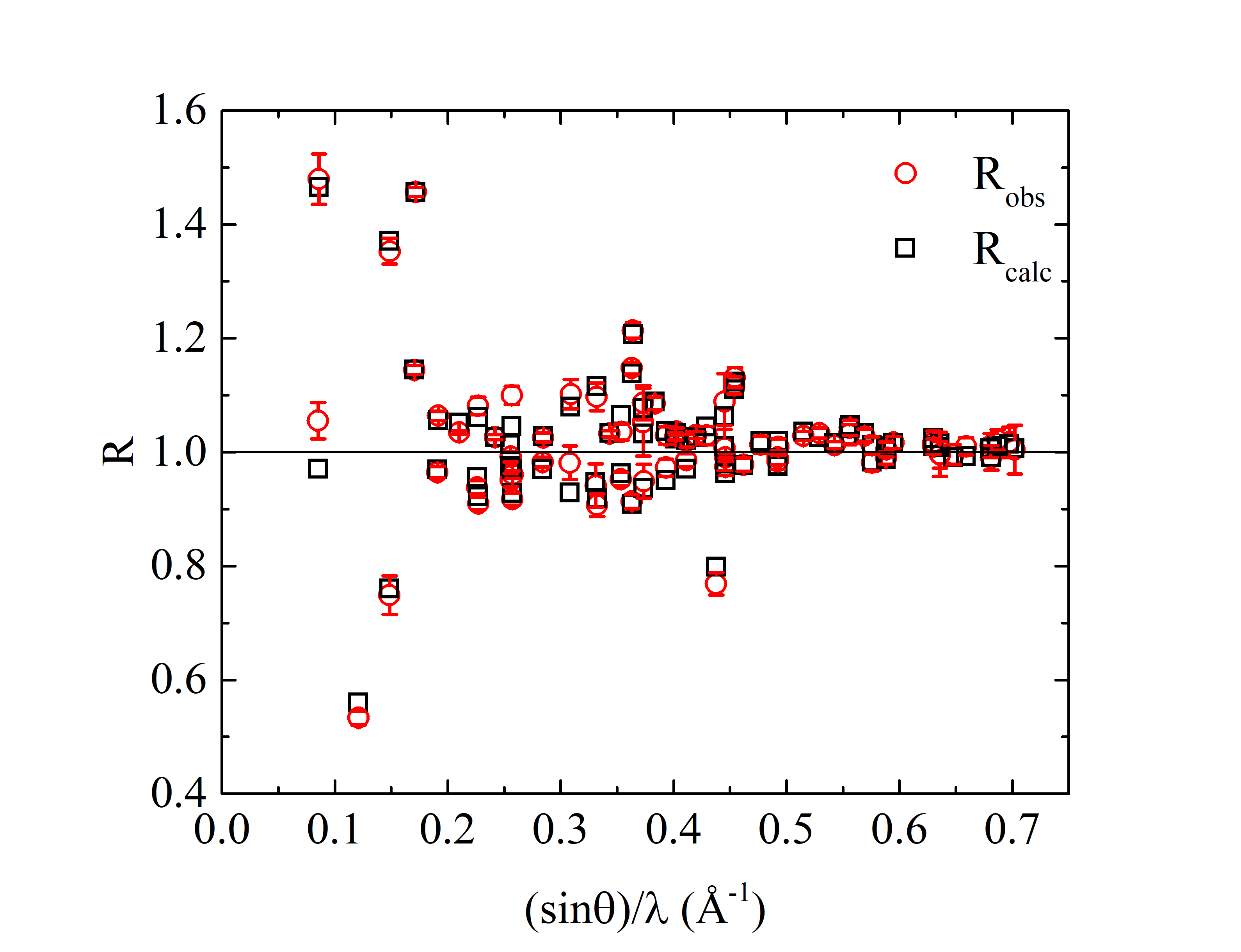}
\par\end{centering}
\caption{\label{fig:Flipping-ratios-measured}Flipping ratios measured at $\unit[1.9]{K}$
and $\unit[9]{T}$ compared with the calculated values using the dipolar
approximation. }

\end{figure}

\begin{table*}
\begin{centering}
\begin{tabular}{c|c|c|ccc|c|c||c|c||c|c}
\hline 
\multirow{3}{*}{} & \multicolumn{2}{c|}{DFT} & \multicolumn{3}{c|}{MAXENT} & \multicolumn{6}{c}{Dipolar approximation}\tabularnewline
\cline{2-12} 
 & \multicolumn{2}{c|}{GGA} & \multirow{2}{*}{$\unit[1]{T}$, $\unit[1.9]{K}$ } & \multirow{2}{*}{$\unit[9]{T}$ , $\unit[1.9]{K}$ } & \multirow{2}{*}{$\unit[9]{T}$ , $\unit[20]{K}$ } & \multicolumn{2}{c}{$\unit[1]{T}$, $\unit[1.9]{K}$ } & \multicolumn{2}{c}{$\unit[9]{T}$ , $\unit[1.9]{K}$ } & \multicolumn{2}{c}{$\unit[9]{T}$ , $\unit[20]{K}$ }\tabularnewline
\cline{7-12} 
 & \multicolumn{2}{c|}{} &  &  &  & \multicolumn{2}{c}{$\chi^{2}=1.27$ } & \multicolumn{2}{c}{$\chi^{2}=1.49$} & \multicolumn{2}{c}{$\chi^{2}=1.02$}\tabularnewline
\hline 
\multirow{2}{*}{Atom, mult.} & $\mu_{\mathrm{S}}$ & \multirow{2}{*}{$\mu$ } & \multirow{2}{*}{$\mu$ } & \multirow{2}{*}{$\mu$ } & \multirow{2}{*}{$\mu$ } & $\mu_{\mathrm{S}}$ & \multirow{2}{*}{$\mu$ } & $\mu_{\mathrm{S}}$ & \multirow{2}{*}{$\mu$ } & $\mu_{\mathrm{S}}$ & \multirow{2}{*}{$\mu$ }\tabularnewline
\cline{2-2} \cline{7-7} \cline{9-9} \cline{11-11} 
 & $\mu_{\mathrm{L}}$  &  &  &  &  & $\mu_{\mathrm{L}}$  &  & $\mu_{\mathrm{L}}$  &  & $\mu_{\mathrm{L}}$  & \tabularnewline
\hline 
\multirow{2}{*}{U1, 1} & -0.657 & \multirow{2}{*}{-0.103} & \multirow{2}{*}{0.11(1)} & \multirow{2}{*}{0.17(1)} & \multirow{2}{*}{0.10(1)} & 0.17(4) & \multirow{2}{*}{0.129(9)} & 0.23(5) & \multirow{2}{*}{0.18(1)} & 0.15(4) & \multirow{2}{*}{0.13(1)}\tabularnewline
\cline{2-2} \cline{7-7} \cline{9-9} \cline{11-11} 
 & 0.554 &  &  &  &  & -0.04(4) &  & -0.05(4) &  & -0.02(4) & \tabularnewline
\cline{2-12} 
\multirow{2}{*}{U2, 3} & -0.820 & \multirow{2}{*}{0.201} & \multirow{2}{*}{0.16(1)} & \multirow{2}{*}{0.22(1)} & \multirow{2}{*}{0.14(1)} & -0.26(2) & \multirow{2}{*}{0.176(6)} & -0.28(3) & \multirow{2}{*}{0.230(8)} & -0.26(2) & \multirow{2}{*}{0.172(6)}\tabularnewline
\cline{2-2} \cline{7-7} \cline{9-9} \cline{11-11} 
 & 1.021 &  &  &  &  & 0.43(2) &  & 0.51(2) &  & 0.43(2) & \tabularnewline
\cline{2-12} 
\multirow{2}{*}{Ru1, 1} & -0.110 & \multirow{2}{*}{-0.115} & \multirow{2}{*}{-} & \multirow{2}{*}{-} & \multirow{2}{*}{-} & 0.04(4) & \multirow{2}{*}{0.01(1)} & 0.00(5) & \multirow{2}{*}{0.00(2)} & 0.00(4) & \multirow{2}{*}{0.00(1)}\tabularnewline
\cline{2-2} \cline{7-7} \cline{9-9} \cline{11-11} 
 & -0.005 &  &  &  &  & -0.03(4) &  & 0.00(4) &  & 0.00(4) & \tabularnewline
\cline{2-12} 
\multirow{2}{*}{Ru2, 6} & 0.119 & \multirow{2}{*}{0.128} & \multirow{2}{*}{-} & \multirow{2}{*}{-} & \multirow{2}{*}{-} & 0.07(2) & \multirow{2}{*}{0.021(5)} & 0.11(2) & \multirow{2}{*}{0.030(6)} & 0.09(1) & \multirow{2}{*}{0.024(4)}\tabularnewline
\cline{2-2} \cline{7-7} \cline{9-9} \cline{11-11} 
 & 0.009 &  &  &  &  & -0.05(2) &  & -0.08(2) &  & -0.07(1) & \tabularnewline
\cline{2-12} 
\multirow{2}{*}{Ge, 6} & 0.009 & \multirow{2}{*}{0.013} & \multirow{2}{*}{-} & \multirow{2}{*}{-} & \multirow{2}{*}{-} & - & \multirow{2}{*}{-} & - & \multirow{2}{*}{-} & - & \multirow{2}{*}{-}\tabularnewline
\cline{2-2} \cline{7-7} \cline{9-9} \cline{11-11} 
 & 0.004 &  &  &  &  & - &  & - &  & - & \tabularnewline
\hline 
\hline 
$\mu_{\mathrm{sum}}$ & \multicolumn{1}{c}{} & 1.231 & 0.85(2) & 1.05(2) & 0.84(2) & \multicolumn{1}{c}{} & \multicolumn{1}{c}{0.79(4)} & \multicolumn{1}{c}{} & \multicolumn{1}{c}{1.02(4)} & \multicolumn{1}{c}{} & 0.79(4)\tabularnewline
$\mu_{\mathrm{bulk}}$ & \multicolumn{1}{c}{} &  & 0.94(1) & 1.25(1) & 0.98(1) & \multicolumn{1}{c}{} & \multicolumn{1}{c}{0.94(1)} & \multicolumn{1}{c}{} & \multicolumn{1}{c}{1.25(1)} & \multicolumn{1}{c}{} & 0.98(1)\tabularnewline
\hline 
\end{tabular}
\par\end{centering}
\caption{\label{tab:Magnetic-moments-calculated}Magnetic moments calculated
from DFT, obtained by MAXENT and from the direct fitting of the flipping
ratios using dipolar approximation in $\unit[1]{T}$ and $\unit[9]{T}$
at $\unit[1.9]{K}$ and $\unit[20]{K}$. All values are in the $\mu_{\mathrm{B}}$.}
\end{table*}

We first discuss the results in the ordered state. The spin and orbital
moments on the U2 site are antiparallel with a dominant orbital contribution
leading to a ratio of $\frac{\mu_{\mathrm{L}}}{\mu_{\mathrm{S}}}=-1.7(2)\,(-1.8(2))$
and parameter $C_{\mathrm{2}}$ reaches a value of $2.5(1)\,(2.3(1))$
at $\unit[1]{T}$ ($\unit[9]{T}$ ).

The orbital component of the magnetic moment on the U1 site is very
small:$\unit[-0.04(4)]{\mu_{\mathrm{B}}}$ and $\unit[-0.05(4)]{\mu_{\mathrm{B}}}$
at $\unit[1]{T}$ and $\unit[9]{T}$, respectively. Together with
the dominating spin component it gives a very unusual value of the
ratio of $\frac{\mu_{\mathrm{L}}}{\mu_{\mathrm{S}}}=-0.2(2)\,(-0.2(2))$
and a very small and negative parameter $C_{\mathrm{2}}=-0.3(3)\,(-0.3(2))$. 

The resulting magnetic moment on the Ru1 position is negligible both
in $\unit[1]{T}$ and $\unit[9]{T}$. On the other hand the Ru2 position
shows a clear induced moment ($\unit[0.021(5)]{\mu_{\mathrm{B}}}$
and $\unit[0.030(6)]{\mu_{\mathrm{B}}}$ for $\unit[1]{T}$ and $\unit[9]{T}$,
respectively). 

The bulk magnetization value $\mu_{\mathrm{bulk}}$ resulting from
the magnetometry measurements increases by a factor 1.3 between $\unit[1]{T}$
($\unit[0.94(1)]{\mu_{\mathrm{B}}}$) and $\unit[9]{T}$ ($\unit[1.25(1)]{\mu_{\mathrm{B}}}$).\cite{Valiska2017}
This scaling is valid within the experimental error, for all the fitted
magnetic moments of the U and Ru ions. The resulting total magnetic
moments per formula unit, $\mu_{\mathrm{sum}}=\mu_{\mathrm{U1}}+3\mu_{\mathrm{U2}}+\mu_{\mathrm{Ru1}}+6\mu_{\mathrm{Ru2}}$
are $\unit[0.79(4)]{\mu_{\mathrm{B}}}$ and $\unit[1.02(4)]{\mu_{\mathrm{B}}}$
for $\unit[1]{T}$ and $\unit[9]{T}$, respectively, smaller than
the bulk magnetization values. The residual moment $\mu_{\mathrm{res}}=\mu_{\mathrm{bulk}}-\mu_{\mathrm{sum}}$
can be attributed to the polarization of the conduction electrons.
Due to the large spatial extent of the conduction electrons their
form factor vanishes around $\frac{\sin\theta}{\lambda}=\unit[0.1]{\textrm{Å}^{-1}}$\cite{Paolasini1993}
and cannot be observed in our neutron diffraction experiments. This
$\mu_{\mathrm{res}}$ value is $\unit[0.15(4)]{\mu_{\mathrm{B}}}$
and $\unit[0.23(4)]{\mu_{\mathrm{B}}}$ for $\unit[1]{T}$ and $\unit[9]{T}$,
respectively reaching $\sim17\%$ of the bulk magnetic moment for
both cases. 

The bulk magnetic moment above the ordering temperature at $\unit[20]{K}$,
induced by the external magnetic field of $\unit[9]{T}$, is $\unit[0.98(1)]{\mu_{\mathrm{B}}}$.
This value is close to the moment in the ordered state at $\unit[1.9]{K}$
in the field of $\unit[1]{T}$, which reaches the value of $\unit[0.94(1)]{\mu_{\mathrm{B}}}$.
The analysis of the flipping ratios obtained in these conditions,
i.e. above $T_{\mathrm{C}}$, at $\unit[20]{K}$ and $\unit[9]{T}$,
gives comparable results to those obtained at $\unit[1.9]{K}$ and
$\unit[1]{T}$, i.e. the U1 and U2 position shows remarkably different
magnetic moments. Results of the fit are summarized in Table \ref{tab:Magnetic-moments-calculated}.
All the components of magnetic moments on all the U and Ru sites are
almost equal within the experimental error. 

\section{Discussion}

Using unpolarized neutron single crystal diffraction methods is not
sufficient to undoubtedly solve the ground-state structure of the
system, as the proposed distortion of the cubic structure is very
small. Comparison of the agreement factors $R_{\mathrm{F}}$ for the
$Im-3m$ and $R-3m$ space group actually favors the cubic structure
both in the ordered state and above $T_{\mathrm{C}}$ (see Table \ref{tab:Structural-parameters}).
However, this may be biased by the fact that the number of inequivalent
reflections is much higher for the rhombohedral structure, and the
internal agreement factor for the equivalent reflections was always
worse for the cubic structure model. The value of the fractional coordinate
$x_{\mathrm{Ru2_{hex}}}$, that is not fixed in the distorted structure,
is close to the value of 0.25 given by symmetry in the cubic description,
for both the measurements at $\unit[20]{K}$ and $\unit[1.9]{K}$.
We can also safely rely on the evidence for the distortion brought
by our previous precision measurement of the thermal expansion.\cite{Valiska2017}
Further proof is brought by the polarized neutron study, which clearly
evidences two uranium sites with drastically different magnetic moments.
Further detailed structural study of this compound, like high resolution
x-ray diffraction, is desired.

The first-principle calculations presented in our previous work\cite{Valiska2017}
and further improved in this paper clearly shows the necessity to
incorporate the spin-orbit interaction while treating the uranium-based
intermetallics. Previous calculations of magnetic moments performed
on the $\mathrm{U_{4}Ru_{7}Ge_{6}}$ omitted the relativistic effects.\cite{Matar2014}
This spin only approach naturally deals with the only one U position
as the lack of spin-orbit interaction does not lower the symmetry.
It gives a total magnetic moment of $\sim\unit[2.72]{\mu_{\mathbf{\mathrm{B}}}/f.u.}$
that overestimates the experimental bulk value. Our results using
GGA give a total magnetic moment of $\sim\unit[1.23]{\mu_{\mathbf{\mathrm{B}}}/f.u.}$,
much closer to the bulk magnetization value ($\sim\unit[1.25]{\mu_{\mathbf{\mathrm{B}}}/f.u.}$
at $\unit[1.9]{K}$ and $\unit[9]{T}$). The spin-orbit interaction
also leads to the appearance of the two distinct U sites (U1 and U2),
in agreement with our proposed rhombohedral distortion. Our DFT calculations
show antiparallel alignment of the spin and orbital magnetic moments
both on the U1 and U2 sites (see Table \ref{tab:Magnetic-moments-calculated}).
This is expected from the third Hund's rule\cite{Aschroft1976} for
systems with less than half-filled shells.The spin moments on the
U1 and U2 sites are parallel. However, the total magnetic moment of
the U1 site is expected to be dominated by its spin component, leading
to a mutually antiparallel alignment of the U1 and U2 total moments
and resulting in a ferrimagnetic structure in the ground state of
$\mathrm{U_{4}Ru_{7}Ge_{6}}$. Such a behavior was observed by neutron
powder diffraction in the case of $\mathrm{UCu_{5}Sn}$ where two
different U sites possess antiparallel collinear magnetic moments
of dramatically different magnitudes of $\unit[2.14(2)]{\mu_{\mathrm{B}}/f.u.}$
on the 2a and $\unit[0.18(4)]{\mu_{\mathrm{B}}/f.u.}$ on the 2c position.\cite{Tran2000}
It is believed to be caused by Kondo screening, acting strongly on
the 2c position. Availability of only unpolarized neutron data prevented
the authors from distinguishing between the spin and orbital components.

Our polarized neutron diffraction results are in good qualitative
agreement with the DFT calculations and gives independent and clear
experimental evidence of the presence of two different U sites, excluding
the cubic $Im-3m$ space group as a proper ground-state structure.
However, while both the maximum entropy calculations and a direct
fit of the flipping ratios clearly show a different density of magnetic
moments on the U1 and U2 sites, they also reveal a positive magnetic
moment on both of them, in disagreement with the DFT calculations
(see Table \ref{tab:Magnetic-moments-calculated}).

Our refined model for fitting of flipping ratios gives a rather large
$C_{\mathrm{2}}\sim2.5$ parameter, irrespective of the applied magnetic
field, for the U2 site. It is much higher than the theoretical values
calculated within the intermediate coupling scheme for the free $\mathrm{U^{3+}}$
(1.64) and $\mathrm{U^{4+}}$ (1.43) ions. These correspond to a ratio
$\frac{\mu_{\mathrm{L}}}{\mu_{\mathrm{S}}}$ of -2.55 and -3.34 for
$\mathrm{U^{3+}}$ and $\mathrm{U^{4+}}$, respectively.\cite{vanderLaan1996}
The U2 site exhibits $\frac{\mu_{\mathrm{L}}}{\mu_{\mathrm{S}}}\sim-1.7$
independent of the applied field. This ratio depends very strongly
on the degree of hybridization between the uranium 5$f$- and transition-metal
$d-$wavefunctions and its decrease from the free ion values means
strengthening of the hybridization.\cite{Lander1991} In that sense
the 5$f$ - wavefunction of the U2 ion strongly hybridizes with the
surrounding Ru 4$d$-wavefunctions as expected from its coordination.
Similar values of the $C_{\mathrm{2}}$ parameter can be found in
the case of antiferromagnetic compounds $\mathrm{UNiGa_{5}}$\cite{Kaneko2003}
($2.45(7)$) and $\mathrm{UGa_{3}}$\cite{Hiess2001} ($2.52(5)$). 

The total U1 magnetic moment, obtained from polarized neutrons is
opposite to the DFT prediction. It also has an extremely small orbital
moment irrespective of the applied magnetic field. This means an almost
quenched orbital moment of the U ion on the U1 position, which is
quite surprising and unexpected. Similar effects was already observed
only in the case of metallic $\alpha$-Uranium, for which the measured
magnetic form factor can be approximated as spin-only and varies significantly
from the one usually observed for the uranium-based compounds.\cite{Maglic1978}
A later theoretical study actually describes $\alpha$-Uranium as
a system where the third Hund's rule is not valid and the orbital
moment can be parallel to the spin moment .\cite{Hjelm1993,Hjelm1994}

Our DFT calculations of $\mathrm{U_{4}Ru_{7}Ge_{6}}$ also predict
non-negligible induced moments on both the Ru sites. They are expected
to be antiparallel to each other with dominating spin components.
A small magnetic moment is also predicted to be present on the Ge
site. In disagreement with the DFT calculations we found only negligible
magnetic moment on the Ru1 site. On the other hand, the Ru2 site exhibits
a significant induced moment dominated by its spin component. These
Ru2 ions form a hexagonal arrangement around the U1 site and a distorted
hexagon around the U2 position. The ratio of the magnetic moment on
the Ru2 site and the U2 site is $\approx0.12$ both at $\unit[1]{T}$
and $\unit[9]{T}$ and $\approx0.16$ for the U1 site. This is another
direct evidence for the hybridization between U 5$f$ and Ru 4$d$
wavefunctions. This resembles the case of URuAl, where the ratio of
the Ru induced magnetic moment to the U moment is even bigger, $0.45(8)$.\cite{Paixao1993}
Parallel alignment of the magnetic moment on the U2 and the Ru2 sites
can be understood by analogy with the mechanism proposed by Brooks
\textit{et al.}\cite{Brooks1989} This was used to describe the parallel
orientation of the moment on the U and the Co sites in $\mathrm{UCoGe}$
observed by XMCD,\cite{Taupin2015}although polarized neutron diffraction
showed an antiparallel arrangement.\cite{Prokes2010} According to
that we would expect our $4d$ spins of Ru to be antiferromagnetically
coupled to the $6d$ spins of U. Thanks to a positive intra-atomic
Hund's rule exchange these U $6d$ spin moments are coupled parallel
to the $5f$ spin moments. Antiferromagnetic coupling of spin and
orbital component on U then result in a final parallel orientation
of the Ru and U moments.

The residual moment in the unit cell $\mu_{\mathrm{res}}$ is estimated
to represent $\sim17\%$ of the bulk moment. A comparably large and
positive ($12\%$) value was observed in antiferromagnetic $\mathrm{UPd_{2}Al_{3}}$
where it was attributed to the possible contribution of the outer
Pd electrons, as there is no observed induced moment on the Pd site
itself.\cite{Paolasini1993} This compound has a similar coordination
of U ions surrounded by hexagons of 4$d$ ions as $\mathrm{U_{4}Ru_{7}Ge_{6}}$
but at a larger distance of $\unit[3.12]{\textrm{Å}}$. An even larger
moment of $\mu_{\mathrm{res}}$ was found in the uranium-based superconductor
UCoGe, where it reaches $54-85\%$ of the bulk magnetic moment\cite{Prokes2010},
depending on the used method.

The fact that we have observed different magnetic moments on U1 and
U2 even in the paramagnetic state at $\unit[20]{K}$ in a $\unit[9]{T}$
applied field clearly shows that the proposed distortion to the rhombohedral
structure with two different U sites can also be induced by an external
magnetic field even above the ordering temperature. The $\frac{\mu_{\mathrm{L}}}{\mu_{\mathrm{S}}}$
ratio remains unchanged relative to the ordered state, showing that
there is no change in the nature of the 5$f$ wavefunctions between
the ferromagnetic and paramagnetic state. Similar results were found
in the case of ferromagnetic superconductor $\mathrm{UGe_{2}}$.\cite{Kernavanois2001}

It has to be noticed, that we assumed only a co-linear (z component)
alignment of the magnetic moments in our study. If we take the $R-3m$
space group as the real ground state structure and assume U1, U2,
Ru1 and Ru2 as the only positions with magnetic moments, then there
are four possible maximal magnetic space groups for the propagation
vector $\left(0,0,0\right)$ (i.e. a ferromagnetic state). Among these
$R-3m^{,}$ is the only one that allows moment on all these four positions.\cite{Aroyo2006}
In the most general case it allows magnetic component out of the z-direction
on the U2 and Ru2 position.

Only very few uranium based systems with two different U sites have
been studied by neutrons. A ferrimagnetic ground-state was confirmed
for the above mentioned $\mathrm{UCu_{5}Sn}$. \cite{Tran2000} Burlet
\textit{et al.}\cite{Burlet1994} describes antiferromagnetic order
of the $\mathrm{U_{4}Cu_{4}P_{7}}$ with two different U sites as
a result of their different valence states. We are unable to make
any conclusions regarding the valence of U ions in the case of $\mathrm{U_{4}Ru_{7}Ge_{6}}$,
since the $\mathrm{U^{3+}}$ and $\mathrm{U^{4+}}$ form factors are
very similar and we also expect rather itinerant nature of the ferromagnetism.
Nevertheless, to our knowledge, $\mathrm{U_{4}Ru_{7}Ge_{6}}$ is so
far the only uranium based ferromagnet that shows a structural distortion
in the ordered state, connected with the appearance of two different,
formerly symmetry equivalent, U sites. 

From the point of view of correct determination of form factors of
dramatically reduced 5$f$-electron magnetic moments and their orbital
and spin components by neutron diffraction $\mathrm{U_{4}Ru_{7}Ge_{6}}$
represents an extremely difficult case for several reasons, mainly:
i) two inequivalent U sites/magnetic moments, ii) lack of reflections
arising from only one of U sites, iii) non-negligible Ru 4$d$-electron
induced moments. Experiments utilizing synchrotron radiation (XMCD,
Compton scattering, resonant X-ray magnetic scattering) are strongly
desired for further progress in understanding the physics of this
unique U intermetallic compound. 

\section{Conclusions }

We have presented experimental evidences of a distortion of the $\mathrm{U_{4}Ru_{7}Ge_{6}}$
cubic lattice in the ferromagnetic state, that was predicted by DFT
calculations and thermal expansion measurements. This transition from
the $Im-3m$ space group to a lower symmetry, most probably described
by $R-3m$ space group, is caused by the dramatic influence of the
spin-orbit interaction on the local symmetry of the U ion site. It
results in the emergence of two crystallographically inequivalent
U sites, U1 and U2, with remarkably different magnetic moments. We
have shown that this effect cannot be observed by usual (unpolarized)
neutron diffraction, but it appears clearly in our polarized neutron
diffraction data. Results of the maximum entropy calculations together
with the model based on dipolar approximation undoubtedly show the
presence of distinct U1 and U2 sites not only in the ground-state
of the compound, but also in the magnetic field (applied along the
easy {[}111{]} direction) induced state in the paramagnetic regime.
Our data suggests a direct connection of the distortion with the magnetic
structure, even when field induced in the paramagnetic state. The
large value of the $C_{\mathrm{2}}\sim2.5$ parameter of the U2 position
suggests an important role of hybridization of $5f$-orbitals with
the surrounding Ru $4d-$wavefunctions and is far from the theoretical
values of the $\mathrm{U^{3+}}$ or $\mathrm{U^{4+}}$ free ions.
Refinement of our data points toward an almost quenched orbital moment
on the U1 site. $\mathrm{U_{4}Ru_{7}Ge_{6}}$ also exhibits a magnetic
moment on the Ru2 position corroborating the strong hybridization
of the 5$f$ U and $4d$ Ru wavefunctions. 
\begin{acknowledgments}
The work was supported within the program of Large Infrastructures
for Research, Experimental Development and Innovation (project No.
LM2015050) and project LTT17019 financed by the Ministry of Education,
Youth and Sports, Czech Republic. This work was supported by the Czech
Science Foundation Grant No. 16-06422S. Experiments were performed
in the Materials Growth and Measurement Laboratory MGML (see: \url{http://mgml.eu/}).
The authors are indebted to Ross H. Colman for reading the manuscript
and providing language corrections.
\end{acknowledgments}

\bibliographystyle{apsrev4-1}
\bibliography{oberdiek-source,UTX}

\begin{thebibliography}{52}%
\makeatletter
\providecommand \@ifxundefined [1]{%
 \@ifx{#1\undefined}
}%
\providecommand \@ifnum [1]{%
 \ifnum #1\expandafter \@firstoftwo
 \else \expandafter \@secondoftwo
 \fi
}%
\providecommand \@ifx [1]{%
 \ifx #1\expandafter \@firstoftwo
 \else \expandafter \@secondoftwo
 \fi
}%
\providecommand \natexlab [1]{#1}%
\providecommand \enquote  [1]{``#1''}%
\providecommand \bibnamefont  [1]{#1}%
\providecommand \bibfnamefont [1]{#1}%
\providecommand \citenamefont [1]{#1}%
\providecommand \href@noop [0]{\@secondoftwo}%
\providecommand \href [0]{\begingroup \@sanitize@url \@href}%
\providecommand \@href[1]{\@@startlink{#1}\@@href}%
\providecommand \@@href[1]{\endgroup#1\@@endlink}%
\providecommand \@sanitize@url [0]{\catcode `\\12\catcode `\$12\catcode
  `\&12\catcode `\#12\catcode `\^12\catcode `\_12\catcode `\%12\relax}%
\providecommand \@@startlink[1]{}%
\providecommand \@@endlink[0]{}%
\providecommand \url  [0]{\begingroup\@sanitize@url \@url }%
\providecommand \@url [1]{\endgroup\@href {#1}{\urlprefix }}%
\providecommand \urlprefix  [0]{URL }%
\providecommand \Eprint [0]{\href }%
\providecommand \doibase [0]{http://dx.doi.org/}%
\providecommand \selectlanguage [0]{\@gobble}%
\providecommand \bibinfo  [0]{\@secondoftwo}%
\providecommand \bibfield  [0]{\@secondoftwo}%
\providecommand \translation [1]{[#1]}%
\providecommand \BibitemOpen [0]{}%
\providecommand \bibitemStop [0]{}%
\providecommand \bibitemNoStop [0]{.\EOS\space}%
\providecommand \EOS [0]{\spacefactor3000\relax}%
\providecommand \BibitemShut  [1]{\csname bibitem#1\endcsname}%
\let\auto@bib@innerbib\@empty
\bibitem [{\citenamefont {Brooks}\ and\ \citenamefont
  {Kelly}(1983)}]{Brooks1983}%
  \BibitemOpen
  \bibfield  {author} {\bibinfo {author} {\bibfnamefont {M.~S.~S.}\
  \bibnamefont {Brooks}}\ and\ \bibinfo {author} {\bibfnamefont {P.~J.}\
  \bibnamefont {Kelly}},\ }\href
  {http://link.aps.org/doi/10.1103/PhysRevLett.51.1708} {\bibfield  {journal}
  {\bibinfo  {journal} {Physical Review Letters}\ }\textbf {\bibinfo {volume}
  {51}},\ \bibinfo {pages} {1708} (\bibinfo {year} {1983})}\BibitemShut
  {NoStop}%
\bibitem [{\citenamefont {Brooks}(1985)}]{Brooks1985}%
  \BibitemOpen
  \bibfield  {author} {\bibinfo {author} {\bibfnamefont {M.~S.~S.}\
  \bibnamefont {Brooks}},\ }\href {\doibase
  https://doi.org/10.1016/0378-4363(85)90170-6} {\bibfield  {journal} {\bibinfo
   {journal} {Physica B+C}\ }\textbf {\bibinfo {volume} {130}},\ \bibinfo
  {pages} {6} (\bibinfo {year} {1985})}\BibitemShut {NoStop}%
\bibitem [{\citenamefont {Eriksson}\ \emph
  {et~al.}(1990{\natexlab{a}})\citenamefont {Eriksson}, \citenamefont
  {Brooks},\ and\ \citenamefont {Johansson}}]{Eriksson1990}%
  \BibitemOpen
  \bibfield  {author} {\bibinfo {author} {\bibfnamefont {O.}~\bibnamefont
  {Eriksson}}, \bibinfo {author} {\bibfnamefont {M.~S.~S.}\ \bibnamefont
  {Brooks}}, \ and\ \bibinfo {author} {\bibfnamefont {B.}~\bibnamefont
  {Johansson}},\ }\href {https://link.aps.org/doi/10.1103/PhysRevB.41.9087}
  {\bibfield  {journal} {\bibinfo  {journal} {Physical Review B}\ }\textbf
  {\bibinfo {volume} {41}},\ \bibinfo {pages} {9087} (\bibinfo {year}
  {1990}{\natexlab{a}})}\BibitemShut {NoStop}%
\bibitem [{\citenamefont {Eriksson}\ \emph
  {et~al.}(1990{\natexlab{b}})\citenamefont {Eriksson}, \citenamefont
  {Brooks},\ and\ \citenamefont {Johansson}}]{Eriksson1990a}%
  \BibitemOpen
  \bibfield  {author} {\bibinfo {author} {\bibfnamefont {O.}~\bibnamefont
  {Eriksson}}, \bibinfo {author} {\bibfnamefont {M.~S.~S.}\ \bibnamefont
  {Brooks}}, \ and\ \bibinfo {author} {\bibfnamefont {B.}~\bibnamefont
  {Johansson}},\ }\href {https://link.aps.org/doi/10.1103/PhysRevB.41.7311}
  {\bibfield  {journal} {\bibinfo  {journal} {Physical Review B}\ }\textbf
  {\bibinfo {volume} {41}},\ \bibinfo {pages} {7311} (\bibinfo {year}
  {1990}{\natexlab{b}})}\BibitemShut {NoStop}%
\bibitem [{\citenamefont {Severin}\ \emph {et~al.}(1991)\citenamefont
  {Severin}, \citenamefont {Nordstr{\"o}m}, \citenamefont {Brooks},\ and\
  \citenamefont {Johansson}}]{Severin1991}%
  \BibitemOpen
  \bibfield  {author} {\bibinfo {author} {\bibfnamefont {L.}~\bibnamefont
  {Severin}}, \bibinfo {author} {\bibfnamefont {L.}~\bibnamefont
  {Nordstr{\"o}m}}, \bibinfo {author} {\bibfnamefont {M.~S.~S.}\ \bibnamefont
  {Brooks}}, \ and\ \bibinfo {author} {\bibfnamefont {B.}~\bibnamefont
  {Johansson}},\ }\href {http://link.aps.org/doi/10.1103/PhysRevB.44.9392}
  {\bibfield  {journal} {\bibinfo  {journal} {Physical Review B}\ }\textbf
  {\bibinfo {volume} {44}},\ \bibinfo {pages} {9392} (\bibinfo {year}
  {1991})}\BibitemShut {NoStop}%
\bibitem [{\citenamefont {Norman}\ and\ \citenamefont
  {Koelling}(1986)}]{Norman1986}%
  \BibitemOpen
  \bibfield  {author} {\bibinfo {author} {\bibfnamefont {M.~R.}\ \bibnamefont
  {Norman}}\ and\ \bibinfo {author} {\bibfnamefont {D.~D.}\ \bibnamefont
  {Koelling}},\ }\href {https://link.aps.org/doi/10.1103/PhysRevB.33.3803}
  {\bibfield  {journal} {\bibinfo  {journal} {Physical Review B}\ }\textbf
  {\bibinfo {volume} {33}},\ \bibinfo {pages} {3803} (\bibinfo {year}
  {1986})}\BibitemShut {NoStop}%
\bibitem [{\citenamefont {Norman}\ \emph {et~al.}(1988)\citenamefont {Norman},
  \citenamefont {Min}, \citenamefont {Oguchi},\ and\ \citenamefont
  {Freeman}}]{Norman1988}%
  \BibitemOpen
  \bibfield  {author} {\bibinfo {author} {\bibfnamefont {M.~R.}\ \bibnamefont
  {Norman}}, \bibinfo {author} {\bibfnamefont {B.~I.}\ \bibnamefont {Min}},
  \bibinfo {author} {\bibfnamefont {T.}~\bibnamefont {Oguchi}}, \ and\ \bibinfo
  {author} {\bibfnamefont {A.~J.}\ \bibnamefont {Freeman}},\ }\href
  {https://link.aps.org/doi/10.1103/PhysRevB.38.6818} {\bibfield  {journal}
  {\bibinfo  {journal} {Physical Review B}\ }\textbf {\bibinfo {volume} {38}},\
  \bibinfo {pages} {6818} (\bibinfo {year} {1988})}\BibitemShut {NoStop}%
\bibitem [{\citenamefont {Wedgewood}\ and\ \citenamefont
  {Kuzneitz}(1972)}]{Wedgewood1972}%
  \BibitemOpen
  \bibfield  {author} {\bibinfo {author} {\bibfnamefont {F.~A.}\ \bibnamefont
  {Wedgewood}}\ and\ \bibinfo {author} {\bibfnamefont {M.}~\bibnamefont
  {Kuzneitz}},\ }\href {http://stacks.iop.org/0022-3719/5/i=21/a=007
  http://iopscience.iop.org/article/10.1088/0022-3719/5/21/007/pdf} {\bibfield
  {journal} {\bibinfo  {journal} {Journal of Physics C: Solid State Physics}\
  }\textbf {\bibinfo {volume} {5}},\ \bibinfo {pages} {3012} (\bibinfo {year}
  {1972})}\BibitemShut {NoStop}%
\bibitem [{\citenamefont {Lander}\ \emph
  {et~al.}(1976{\natexlab{a}})\citenamefont {Lander}, \citenamefont {Faber},
  \citenamefont {Freeman},\ and\ \citenamefont {Desclaux}}]{Lander1976a}%
  \BibitemOpen
  \bibfield  {author} {\bibinfo {author} {\bibfnamefont {G.~H.}\ \bibnamefont
  {Lander}}, \bibinfo {author} {\bibfnamefont {J.}~\bibnamefont {Faber}},
  \bibinfo {author} {\bibfnamefont {A.~J.}\ \bibnamefont {Freeman}}, \ and\
  \bibinfo {author} {\bibfnamefont {J.~P.}\ \bibnamefont {Desclaux}},\ }\href
  {https://link.aps.org/doi/10.1103/PhysRevB.13.1177} {\bibfield  {journal}
  {\bibinfo  {journal} {Physical Review B}\ }\textbf {\bibinfo {volume} {13}},\
  \bibinfo {pages} {1177} (\bibinfo {year} {1976}{\natexlab{a}})}\BibitemShut
  {NoStop}%
\bibitem [{\citenamefont {Faber}\ and\ \citenamefont
  {Lander}(1976)}]{Faber1976}%
  \BibitemOpen
  \bibfield  {author} {\bibinfo {author} {\bibfnamefont {J.}~\bibnamefont
  {Faber}}\ and\ \bibinfo {author} {\bibfnamefont {G.~H.}\ \bibnamefont
  {Lander}},\ }\href {https://link.aps.org/doi/10.1103/PhysRevB.14.1151}
  {\bibfield  {journal} {\bibinfo  {journal} {Physical Review B}\ }\textbf
  {\bibinfo {volume} {14}},\ \bibinfo {pages} {1151} (\bibinfo {year}
  {1976})}\BibitemShut {NoStop}%
\bibitem [{\citenamefont {Lander}\ \emph
  {et~al.}(1976{\natexlab{b}})\citenamefont {Lander}, \citenamefont {Mueller},
  \citenamefont {Sparlin},\ and\ \citenamefont {Vogt}}]{Lander1976b}%
  \BibitemOpen
  \bibfield  {author} {\bibinfo {author} {\bibfnamefont {G.~H.}\ \bibnamefont
  {Lander}}, \bibinfo {author} {\bibfnamefont {M.~H.}\ \bibnamefont {Mueller}},
  \bibinfo {author} {\bibfnamefont {D.~M.}\ \bibnamefont {Sparlin}}, \ and\
  \bibinfo {author} {\bibfnamefont {O.}~\bibnamefont {Vogt}},\ }\href
  {https://link.aps.org/doi/10.1103/PhysRevB.14.5035} {\bibfield  {journal}
  {\bibinfo  {journal} {Physical Review B}\ }\textbf {\bibinfo {volume} {14}},\
  \bibinfo {pages} {5035} (\bibinfo {year} {1976}{\natexlab{b}})}\BibitemShut
  {NoStop}%
\bibitem [{\citenamefont {Delapalme}\ \emph {et~al.}(1978)\citenamefont
  {Delapalme}, \citenamefont {Lander},\ and\ \citenamefont
  {Brown}}]{Delapalme1978}%
  \BibitemOpen
  \bibfield  {author} {\bibinfo {author} {\bibfnamefont {A.}~\bibnamefont
  {Delapalme}}, \bibinfo {author} {\bibfnamefont {G.~H.}\ \bibnamefont
  {Lander}}, \ and\ \bibinfo {author} {\bibfnamefont {P.~J.}\ \bibnamefont
  {Brown}},\ }\href {http://stacks.iop.org/0022-3719/11/i=7/a=033
  http://iopscience.iop.org/article/10.1088/0022-3719/11/7/033/pdf} {\bibfield
  {journal} {\bibinfo  {journal} {Journal of Physics C: Solid State Physics}\
  }\textbf {\bibinfo {volume} {11}},\ \bibinfo {pages} {1441} (\bibinfo {year}
  {1978})}\BibitemShut {NoStop}%
\bibitem [{\citenamefont {Koelling}\ \emph {et~al.}(1985)\citenamefont
  {Koelling}, \citenamefont {Dunlap},\ and\ \citenamefont
  {Crabtree}}]{Koelling1985}%
  \BibitemOpen
  \bibfield  {author} {\bibinfo {author} {\bibfnamefont {D.~D.}\ \bibnamefont
  {Koelling}}, \bibinfo {author} {\bibfnamefont {B.~D.}\ \bibnamefont
  {Dunlap}}, \ and\ \bibinfo {author} {\bibfnamefont {G.~W.}\ \bibnamefont
  {Crabtree}},\ }\href {http://link.aps.org/doi/10.1103/PhysRevB.31.4966}
  {\bibfield  {journal} {\bibinfo  {journal} {Physical Review B}\ }\textbf
  {\bibinfo {volume} {31}},\ \bibinfo {pages} {4966} (\bibinfo {year}
  {1985})}\BibitemShut {NoStop}%
\bibitem [{\citenamefont {Eriksson}\ \emph {et~al.}(1989)\citenamefont
  {Eriksson}, \citenamefont {Johansson}, \citenamefont {Brooks},\ and\
  \citenamefont {Skriver}}]{Eriksson1989}%
  \BibitemOpen
  \bibfield  {author} {\bibinfo {author} {\bibfnamefont {O.}~\bibnamefont
  {Eriksson}}, \bibinfo {author} {\bibfnamefont {B.}~\bibnamefont {Johansson}},
  \bibinfo {author} {\bibfnamefont {M.~S.~S.}\ \bibnamefont {Brooks}}, \ and\
  \bibinfo {author} {\bibfnamefont {H.~L.}\ \bibnamefont {Skriver}},\ }\href
  {https://link.aps.org/doi/10.1103/PhysRevB.40.9508} {\bibfield  {journal}
  {\bibinfo  {journal} {Physical Review B}\ }\textbf {\bibinfo {volume} {40}},\
  \bibinfo {pages} {9508} (\bibinfo {year} {1989})}\BibitemShut {NoStop}%
\bibitem [{\citenamefont {Sechovsk{\'y}}\ \emph {et~al.}(1980)\citenamefont
  {Sechovsk{\'y}}, \citenamefont {Smetana}, \citenamefont {Hilscher},
  \citenamefont {Gratz},\ and\ \citenamefont {Sassik}}]{Sechovsky1980}%
  \BibitemOpen
  \bibfield  {author} {\bibinfo {author} {\bibfnamefont {V.}~\bibnamefont
  {Sechovsk{\'y}}}, \bibinfo {author} {\bibfnamefont {Z.}~\bibnamefont
  {Smetana}}, \bibinfo {author} {\bibfnamefont {G.}~\bibnamefont {Hilscher}},
  \bibinfo {author} {\bibfnamefont {E.}~\bibnamefont {Gratz}}, \ and\ \bibinfo
  {author} {\bibfnamefont {H.}~\bibnamefont {Sassik}},\ }\href {\doibase
  http://dx.doi.org/10.1016/0378-4363(80)90173-4} {\bibfield  {journal}
  {\bibinfo  {journal} {Physica B+C}\ }\textbf {\bibinfo {volume} {102}},\
  \bibinfo {pages} {277} (\bibinfo {year} {1980})}\BibitemShut {NoStop}%
\bibitem [{\citenamefont {Fournier}\ \emph {et~al.}(1986)\citenamefont
  {Fournier}, \citenamefont {Boeuf}, \citenamefont {Frings}, \citenamefont
  {Bonnet}, \citenamefont {Boucherle}, \citenamefont {Delapalme},\ and\
  \citenamefont {Menovsky}}]{Fournier1986}%
  \BibitemOpen
  \bibfield  {author} {\bibinfo {author} {\bibfnamefont {J.~M.}\ \bibnamefont
  {Fournier}}, \bibinfo {author} {\bibfnamefont {A.}~\bibnamefont {Boeuf}},
  \bibinfo {author} {\bibfnamefont {P.}~\bibnamefont {Frings}}, \bibinfo
  {author} {\bibfnamefont {M.}~\bibnamefont {Bonnet}}, \bibinfo {author}
  {\bibfnamefont {J.~v.}\ \bibnamefont {Boucherle}}, \bibinfo {author}
  {\bibfnamefont {A.}~\bibnamefont {Delapalme}}, \ and\ \bibinfo {author}
  {\bibfnamefont {A.}~\bibnamefont {Menovsky}},\ }\href {\doibase
  http://dx.doi.org/10.1016/0022-5088(86)90537-0} {\bibfield  {journal}
  {\bibinfo  {journal} {Journal of the Less Common Metals}\ }\textbf {\bibinfo
  {volume} {121}},\ \bibinfo {pages} {249} (\bibinfo {year}
  {1986})}\BibitemShut {NoStop}%
\bibitem [{\citenamefont {Wulff}\ \emph {et~al.}(1989)\citenamefont {Wulff},
  \citenamefont {Lander}, \citenamefont {Lebech},\ and\ \citenamefont
  {Delapalme}}]{Wulff1989}%
  \BibitemOpen
  \bibfield  {author} {\bibinfo {author} {\bibfnamefont {M.}~\bibnamefont
  {Wulff}}, \bibinfo {author} {\bibfnamefont {G.~H.}\ \bibnamefont {Lander}},
  \bibinfo {author} {\bibfnamefont {B.}~\bibnamefont {Lebech}}, \ and\ \bibinfo
  {author} {\bibfnamefont {A.}~\bibnamefont {Delapalme}},\ }\href
  {https://link.aps.org/doi/10.1103/PhysRevB.39.4719} {\bibfield  {journal}
  {\bibinfo  {journal} {Physical Review B}\ }\textbf {\bibinfo {volume} {39}},\
  \bibinfo {pages} {4719} (\bibinfo {year} {1989})}\BibitemShut {NoStop}%
\bibitem [{\citenamefont {Lebech}\ \emph {et~al.}(1991)\citenamefont {Lebech},
  \citenamefont {Wulff},\ and\ \citenamefont {Lander}}]{Lebech1991}%
  \BibitemOpen
  \bibfield  {author} {\bibinfo {author} {\bibfnamefont {B.}~\bibnamefont
  {Lebech}}, \bibinfo {author} {\bibfnamefont {M.}~\bibnamefont {Wulff}}, \
  and\ \bibinfo {author} {\bibfnamefont {G.~H.}\ \bibnamefont {Lander}},\
  }\href {\doibase doi:http://dx.doi.org/10.1063/1.347864} {\bibfield
  {journal} {\bibinfo  {journal} {Journal of Applied Physics}\ }\textbf
  {\bibinfo {volume} {69}},\ \bibinfo {pages} {5891} (\bibinfo {year}
  {1991})}\BibitemShut {NoStop}%
\bibitem [{\citenamefont {Brooks}\ \emph {et~al.}(1988)\citenamefont {Brooks},
  \citenamefont {Eriksson}, \citenamefont {Johansson}, \citenamefont {Franse},\
  and\ \citenamefont {Frings}}]{Brooks1988}%
  \BibitemOpen
  \bibfield  {author} {\bibinfo {author} {\bibfnamefont {M.~S.~S.}\
  \bibnamefont {Brooks}}, \bibinfo {author} {\bibfnamefont {O.}~\bibnamefont
  {Eriksson}}, \bibinfo {author} {\bibfnamefont {B.}~\bibnamefont {Johansson}},
  \bibinfo {author} {\bibfnamefont {J.~J.~M.}\ \bibnamefont {Franse}}, \ and\
  \bibinfo {author} {\bibfnamefont {P.~H.}\ \bibnamefont {Frings}},\ }\href
  {http://stacks.iop.org/0305-4608/18/i=3/a=003
  http://iopscience.iop.org/article/10.1088/0305-4608/18/3/003/pdf} {\bibfield
  {journal} {\bibinfo  {journal} {Journal of Physics F: Metal Physics}\
  }\textbf {\bibinfo {volume} {18}},\ \bibinfo {pages} {L33} (\bibinfo {year}
  {1988})}\BibitemShut {NoStop}%
\bibitem [{\citenamefont {Menovsky}(1988)}]{Menovsky1988}%
  \BibitemOpen
  \bibfield  {author} {\bibinfo {author} {\bibfnamefont {A.~A.}\ \bibnamefont
  {Menovsky}},\ }\href {\doibase
  http://dx.doi.org/10.1016/0304-8853(88)90510-0} {\bibfield  {journal}
  {\bibinfo  {journal} {Journal of Magnetism and Magnetic Materials}\ }\textbf
  {\bibinfo {volume} {76-77}},\ \bibinfo {pages} {631} (\bibinfo {year}
  {1988})}\BibitemShut {NoStop}%
\bibitem [{\citenamefont {Lloret}\ \emph {et~al.}(1987)\citenamefont {Lloret},
  \citenamefont {Buffat}, \citenamefont {Chevalier},\ and\ \citenamefont
  {Etourneau}}]{Lloret1987}%
  \BibitemOpen
  \bibfield  {author} {\bibinfo {author} {\bibfnamefont {B.}~\bibnamefont
  {Lloret}}, \bibinfo {author} {\bibfnamefont {B.}~\bibnamefont {Buffat}},
  \bibinfo {author} {\bibfnamefont {B.}~\bibnamefont {Chevalier}}, \ and\
  \bibinfo {author} {\bibfnamefont {J.}~\bibnamefont {Etourneau}},\ }\href
  {\doibase http://dx.doi.org/10.1016/0304-8853(87)90236-8} {\bibfield
  {journal} {\bibinfo  {journal} {Journal of Magnetism and Magnetic Materials}\
  }\textbf {\bibinfo {volume} {67}},\ \bibinfo {pages} {232} (\bibinfo {year}
  {1987})}\BibitemShut {NoStop}%
\bibitem [{\citenamefont {Mentink}\ \emph {et~al.}(1991)\citenamefont
  {Mentink}, \citenamefont {Nieuwenhuys}, \citenamefont {Menovsky},\ and\
  \citenamefont {Mydosh}}]{Mentink1991}%
  \BibitemOpen
  \bibfield  {author} {\bibinfo {author} {\bibfnamefont {S.~A.~M.}\
  \bibnamefont {Mentink}}, \bibinfo {author} {\bibfnamefont {G.~J.}\
  \bibnamefont {Nieuwenhuys}}, \bibinfo {author} {\bibfnamefont {A.~A.}\
  \bibnamefont {Menovsky}}, \ and\ \bibinfo {author} {\bibfnamefont {J.~A.}\
  \bibnamefont {Mydosh}},\ }\href {\doibase
  doi:http://dx.doi.org/10.1063/1.348942} {\bibfield  {journal} {\bibinfo
  {journal} {Journal of Applied Physics}\ }\textbf {\bibinfo {volume} {69}},\
  \bibinfo {pages} {5484} (\bibinfo {year} {1991})}\BibitemShut {NoStop}%
\bibitem [{\citenamefont {Vali\v{s}ka}\ \emph {et~al.}(2017)\citenamefont
  {Vali\v{s}ka}, \citenamefont {Divi\v{s}},\ and\ \citenamefont
  {Sechovsk\'{y}}}]{Valiska2017}%
  \BibitemOpen
  \bibfield  {author} {\bibinfo {author} {\bibfnamefont {M.}~\bibnamefont
  {Vali\v{s}ka}}, \bibinfo {author} {\bibfnamefont {M.}~\bibnamefont
  {Divi\v{s}}}, \ and\ \bibinfo {author} {\bibfnamefont {V.}~\bibnamefont
  {Sechovsk\'{y}}},\ }\href
  {https://link.aps.org/doi/10.1103/PhysRevB.95.085142} {\bibfield  {journal}
  {\bibinfo  {journal} {Physical Review B}\ }\textbf {\bibinfo {volume} {95}},\
  \bibinfo {pages} {085142} (\bibinfo {year} {2017})}\BibitemShut {NoStop}%
\bibitem [{\citenamefont {Vali\v{s}ka}\ \emph {et~al.}(2016)\citenamefont
  {Vali\v{s}ka}, \citenamefont {Fabelo~Rosa}, \citenamefont {Klicpera},
  \citenamefont {Sechovsk\'{y}},\ and\ \citenamefont {Stunault}}]{Valiska2016}%
  \BibitemOpen
  \bibfield  {author} {\bibinfo {author} {\bibfnamefont {M.}~\bibnamefont
  {Vali\v{s}ka}}, \bibinfo {author} {\bibfnamefont {O.~R.}\ \bibnamefont
  {Fabelo~Rosa}}, \bibinfo {author} {\bibfnamefont {M.}~\bibnamefont
  {Klicpera}}, \bibinfo {author} {\bibfnamefont {V.}~\bibnamefont
  {Sechovsk\'{y}}}, \ and\ \bibinfo {author} {\bibfnamefont {A.}~\bibnamefont
  {Stunault}},\ }\href {\doibase 10.5291/ILL-DATA.5-51-516} {\enquote {\bibinfo
  {title} {Structure change and spin density distribution in unique
  magnetically low-anisotropic compound
  ${\mathrm{u}}_{4}{\mathrm{ru}}_{7}{\mathrm{ge}}_{6}$.}}\ } (\bibinfo {year}
  {2016})\BibitemShut {NoStop}%
\bibitem [{\citenamefont {Aroyo~Mois}\ \emph {et~al.}(2006)\citenamefont
  {Aroyo~Mois}, \citenamefont {Perez-Mato~Juan}, \citenamefont {Capillas},
  \citenamefont {Kroumova}, \citenamefont {Ivantchev}, \citenamefont
  {Madariaga}, \citenamefont {Kirov},\ and\ \citenamefont
  {Wondratschek}}]{Aroyo2006}%
  \BibitemOpen
  \bibfield  {author} {\bibinfo {author} {\bibfnamefont {I.}~\bibnamefont
  {Aroyo~Mois}}, \bibinfo {author} {\bibfnamefont {M.}~\bibnamefont
  {Perez-Mato~Juan}}, \bibinfo {author} {\bibfnamefont {C.}~\bibnamefont
  {Capillas}}, \bibinfo {author} {\bibfnamefont {E.}~\bibnamefont {Kroumova}},
  \bibinfo {author} {\bibfnamefont {S.}~\bibnamefont {Ivantchev}}, \bibinfo
  {author} {\bibfnamefont {G.}~\bibnamefont {Madariaga}}, \bibinfo {author}
  {\bibfnamefont {A.}~\bibnamefont {Kirov}}, \ and\ \bibinfo {author}
  {\bibfnamefont {H.}~\bibnamefont {Wondratschek}},\ }\href {\doibase
  10.1524/zkri.2006.221.1.15} {\enquote {\bibinfo {title} {Bilbao
  crystallographic server: I. databases and crystallographic computing
  programs},}\ } (\bibinfo {year} {2006})\BibitemShut {NoStop}%
\bibitem [{\citenamefont {Roisnel}\ and\ \citenamefont
  {Rodriguez-Carvajal}()}]{Roisnel2000}%
  \BibitemOpen
  \bibfield  {author} {\bibinfo {author} {\bibfnamefont {T.}~\bibnamefont
  {Roisnel}}\ and\ \bibinfo {author} {\bibfnamefont {J.}~\bibnamefont
  {Rodriguez-Carvajal}},\ }in\ \href
  {http://books.google.cz/books?id=j01QAAAAYAAJ} {\emph {\bibinfo {booktitle}
  {EPDIC 7 - Seventh European Powder Diffraction Conference}}},\ \bibinfo
  {editor} {edited by\ \bibinfo {editor} {\bibfnamefont {R.}~\bibnamefont
  {Delhez}}\ and\ \bibinfo {editor} {\bibfnamefont {E.}~\bibnamefont
  {Mittemeijer}}}\ (\bibinfo  {publisher} {Trans Tech
  Publications})\BibitemShut {NoStop}%
\bibitem [{\citenamefont {Rodriguez-Carvajal}(1993)}]{Rodriguez-Carvajal1993}%
  \BibitemOpen
  \bibfield  {author} {\bibinfo {author} {\bibfnamefont {J.}~\bibnamefont
  {Rodriguez-Carvajal}},\ }\href {\doibase
  http://dx.doi.org/10.1016/0921-4526(93)90108-I} {\bibfield  {journal}
  {\bibinfo  {journal} {Physica B: Condensed Matter}\ }\textbf {\bibinfo
  {volume} {192}},\ \bibinfo {pages} {55} (\bibinfo {year} {1993})}\BibitemShut
  {NoStop}%
\bibitem [{\citenamefont {Koepernik}\ and\ \citenamefont
  {Eschrig}(1999)}]{Koepernik}%
  \BibitemOpen
  \bibfield  {author} {\bibinfo {author} {\bibfnamefont {K.}~\bibnamefont
  {Koepernik}}\ and\ \bibinfo {author} {\bibfnamefont {H.}~\bibnamefont
  {Eschrig}},\ }\href {http://link.aps.org/doi/10.1103/PhysRevB.59.1743}
  {\bibfield  {journal} {\bibinfo  {journal} {Physical Review B}\ }\textbf
  {\bibinfo {volume} {59}},\ \bibinfo {pages} {1743} (\bibinfo {year}
  {1999})}\BibitemShut {NoStop}%
\bibitem [{\citenamefont {Perdew}\ and\ \citenamefont
  {Wang}(1992)}]{Perdew1992}%
  \BibitemOpen
  \bibfield  {author} {\bibinfo {author} {\bibfnamefont {J.~P.}\ \bibnamefont
  {Perdew}}\ and\ \bibinfo {author} {\bibfnamefont {Y.}~\bibnamefont {Wang}},\
  }\href {http://link.aps.org/doi/10.1103/PhysRevB.45.13244} {\bibfield
  {journal} {\bibinfo  {journal} {Physical Review B}\ }\textbf {\bibinfo
  {volume} {45}},\ \bibinfo {pages} {13244} (\bibinfo {year}
  {1992})}\BibitemShut {NoStop}%
\bibitem [{\citenamefont {Perdew}\ \emph {et~al.}(1996)\citenamefont {Perdew},
  \citenamefont {Burke},\ and\ \citenamefont {Ernzerhof}}]{Perdew1996}%
  \BibitemOpen
  \bibfield  {author} {\bibinfo {author} {\bibfnamefont {J.~P.}\ \bibnamefont
  {Perdew}}, \bibinfo {author} {\bibfnamefont {K.}~\bibnamefont {Burke}}, \
  and\ \bibinfo {author} {\bibfnamefont {M.}~\bibnamefont {Ernzerhof}},\ }\href
  {http://link.aps.org/doi/10.1103/PhysRevLett.77.3865} {\bibfield  {journal}
  {\bibinfo  {journal} {Physical Review Letters}\ }\textbf {\bibinfo {volume}
  {77}},\ \bibinfo {pages} {3865} (\bibinfo {year} {1996})}\BibitemShut
  {NoStop}%
\bibitem [{\citenamefont {Momma}\ \emph {et~al.}(2013)\citenamefont {Momma},
  \citenamefont {Ikeda}, \citenamefont {Belik},\ and\ \citenamefont
  {Izumi}}]{Momma2013}%
  \BibitemOpen
  \bibfield  {author} {\bibinfo {author} {\bibfnamefont {K.}~\bibnamefont
  {Momma}}, \bibinfo {author} {\bibfnamefont {T.}~\bibnamefont {Ikeda}},
  \bibinfo {author} {\bibfnamefont {A.~A.}\ \bibnamefont {Belik}}, \ and\
  \bibinfo {author} {\bibfnamefont {F.}~\bibnamefont {Izumi}},\ }\href
  {\doibase 10.1017/s088571561300002x} {\bibfield  {journal} {\bibinfo
  {journal} {Powder Diffraction}\ }\textbf {\bibinfo {volume} {28}},\ \bibinfo
  {pages} {184} (\bibinfo {year} {2013})}\BibitemShut {NoStop}%
\bibitem [{\citenamefont {Momma}\ and\ \citenamefont
  {Izumi}(2011)}]{Momma2011}%
  \BibitemOpen
  \bibfield  {author} {\bibinfo {author} {\bibfnamefont {K.}~\bibnamefont
  {Momma}}\ and\ \bibinfo {author} {\bibfnamefont {F.}~\bibnamefont {Izumi}},\
  }\href {\doibase doi:10.1107/S0021889811038970} {\bibfield  {journal}
  {\bibinfo  {journal} {Journal of Applied Crystallography}\ }\textbf {\bibinfo
  {volume} {44}},\ \bibinfo {pages} {1272} (\bibinfo {year}
  {2011})}\BibitemShut {NoStop}%
\bibitem [{\citenamefont {Slater}(1964)}]{Slater1964}%
  \BibitemOpen
  \bibfield  {author} {\bibinfo {author} {\bibfnamefont {J.~C.}\ \bibnamefont
  {Slater}},\ }\href {\doibase 10.1063/1.1725697} {\bibfield  {journal}
  {\bibinfo  {journal} {The Journal of Chemical Physics}\ }\textbf {\bibinfo
  {volume} {41}},\ \bibinfo {pages} {3199} (\bibinfo {year}
  {1964})}\BibitemShut {NoStop}%
\bibitem [{\citenamefont {Brown}(2006)}]{Brown}%
  \BibitemOpen
  \bibfield  {author} {\bibinfo {author} {\bibfnamefont {P.~J.}\ \bibnamefont
  {Brown}},\ }\enquote {\bibinfo {title} {Magnetic form factors},}\ in\ \href
  {\doibase 10.1107/97809553602060000103} {\emph {\bibinfo {booktitle}
  {International Tables for Crystallography}}},\ Vol.~\bibinfo {volume} {C},\
  \bibinfo {editor} {edited by\ \bibinfo {editor} {\bibfnamefont
  {E.}~\bibnamefont {Prince}}}\ (\bibinfo  {publisher} {Wiley},\ \bibinfo
  {year} {2006})\ \bibinfo {type} {Book section}\ \bibinfo {chapter} {4.4},
  pp.\ \bibinfo {pages} {454--461}\BibitemShut {NoStop}%
\bibitem [{\citenamefont {Freeman}\ \emph {et~al.}(1976)\citenamefont
  {Freeman}, \citenamefont {Desclaux}, \citenamefont {Lander},\ and\
  \citenamefont {Faber}}]{Freeman1976}%
  \BibitemOpen
  \bibfield  {author} {\bibinfo {author} {\bibfnamefont {A.~J.}\ \bibnamefont
  {Freeman}}, \bibinfo {author} {\bibfnamefont {J.~P.}\ \bibnamefont
  {Desclaux}}, \bibinfo {author} {\bibfnamefont {G.~H.}\ \bibnamefont
  {Lander}}, \ and\ \bibinfo {author} {\bibfnamefont {J.}~\bibnamefont
  {Faber}},\ }\href {https://link.aps.org/doi/10.1103/PhysRevB.13.1168}
  {\bibfield  {journal} {\bibinfo  {journal} {Physical Review B}\ }\textbf
  {\bibinfo {volume} {13}},\ \bibinfo {pages} {1168} (\bibinfo {year}
  {1976})}\BibitemShut {NoStop}%
\bibitem [{\citenamefont {Paolasini}\ \emph {et~al.}(1993)\citenamefont
  {Paolasini}, \citenamefont {Paixao}, \citenamefont {Lander}, \citenamefont
  {Delapalme}, \citenamefont {Sato},\ and\ \citenamefont
  {Komatsubara}}]{Paolasini1993}%
  \BibitemOpen
  \bibfield  {author} {\bibinfo {author} {\bibfnamefont {L.}~\bibnamefont
  {Paolasini}}, \bibinfo {author} {\bibfnamefont {J.}~\bibnamefont {Paixao}},
  \bibinfo {author} {\bibfnamefont {G.~H.}\ \bibnamefont {Lander}}, \bibinfo
  {author} {\bibfnamefont {A.}~\bibnamefont {Delapalme}}, \bibinfo {author}
  {\bibfnamefont {N.}~\bibnamefont {Sato}}, \ and\ \bibinfo {author}
  {\bibfnamefont {T.}~\bibnamefont {Komatsubara}},\ }\href
  {http://stacks.iop.org/0953-8984/5/i=47/a=015
  http://iopscience.iop.org/article/10.1088/0953-8984/5/47/015/pdf} {\bibfield
  {journal} {\bibinfo  {journal} {Journal of Physics: Condensed Matter}\
  }\textbf {\bibinfo {volume} {5}},\ \bibinfo {pages} {8905} (\bibinfo {year}
  {1993})}\BibitemShut {NoStop}%
\bibitem [{\citenamefont {Matar}\ \emph {et~al.}(2014)\citenamefont {Matar},
  \citenamefont {Chevalier},\ and\ \citenamefont {P{\"o}ttgen}}]{Matar2014}%
  \BibitemOpen
  \bibfield  {author} {\bibinfo {author} {\bibfnamefont {S.~F.}\ \bibnamefont
  {Matar}}, \bibinfo {author} {\bibfnamefont {B.}~\bibnamefont {Chevalier}}, \
  and\ \bibinfo {author} {\bibfnamefont {R.}~\bibnamefont {P{\"o}ttgen}},\
  }\href {\doibase http://dx.doi.org/10.1016/j.solidstatesciences.2013.10.008}
  {\bibfield  {journal} {\bibinfo  {journal} {Solid State Sciences}\ }\textbf
  {\bibinfo {volume} {27}},\ \bibinfo {pages} {5} (\bibinfo {year}
  {2014})}\BibitemShut {NoStop}%
\bibitem [{\citenamefont {Ashcroft}\ and\ \citenamefont
  {Mermin}(1976)}]{Aschroft1976}%
  \BibitemOpen
  \bibfield  {author} {\bibinfo {author} {\bibfnamefont {N.}~\bibnamefont
  {Ashcroft}}\ and\ \bibinfo {author} {\bibfnamefont {N.}~\bibnamefont
  {Mermin}},\ }\enquote {\bibinfo {title} {Solid state physics},}\ \ (\bibinfo
  {publisher} {Saunders College},\ \bibinfo {year} {1976})\ \bibinfo {type}
  {Book section}~\bibinfo {chapter} {31}\BibitemShut {NoStop}%
\bibitem [{\citenamefont {Tran}\ \emph {et~al.}(2000)\citenamefont {Tran},
  \citenamefont {Troc}, \citenamefont {Andr\'{e}}, \citenamefont {Bour\'{e}e},\
  and\ \citenamefont {Kolenda}}]{Tran2000}%
  \BibitemOpen
  \bibfield  {author} {\bibinfo {author} {\bibfnamefont {V.~H.}\ \bibnamefont
  {Tran}}, \bibinfo {author} {\bibfnamefont {R.}~\bibnamefont {Troc}}, \bibinfo
  {author} {\bibfnamefont {G.}~\bibnamefont {Andr\'{e}}}, \bibinfo {author}
  {\bibfnamefont {F.}~\bibnamefont {Bour\'{e}e}}, \ and\ \bibinfo {author}
  {\bibfnamefont {M.}~\bibnamefont {Kolenda}},\ }\href
  {http://stacks.iop.org/0953-8984/12/i=27/a=306
  http://iopscience.iop.org/article/10.1088/0953-8984/12/27/306/pdf} {\bibfield
   {journal} {\bibinfo  {journal} {Journal of Physics: Condensed Matter}\
  }\textbf {\bibinfo {volume} {12}},\ \bibinfo {pages} {5879} (\bibinfo {year}
  {2000})}\BibitemShut {NoStop}%
\bibitem [{\citenamefont {van~der Laan}\ and\ \citenamefont
  {Thole}(1996)}]{vanderLaan1996}%
  \BibitemOpen
  \bibfield  {author} {\bibinfo {author} {\bibfnamefont {G.}~\bibnamefont
  {van~der Laan}}\ and\ \bibinfo {author} {\bibfnamefont {B.~T.}\ \bibnamefont
  {Thole}},\ }\href {https://link.aps.org/doi/10.1103/PhysRevB.53.14458}
  {\bibfield  {journal} {\bibinfo  {journal} {Physical Review B}\ }\textbf
  {\bibinfo {volume} {53}},\ \bibinfo {pages} {14458} (\bibinfo {year}
  {1996})}\BibitemShut {NoStop}%
\bibitem [{\citenamefont {Lander}\ \emph {et~al.}(1991)\citenamefont {Lander},
  \citenamefont {Brooks},\ and\ \citenamefont {Johansson}}]{Lander1991}%
  \BibitemOpen
  \bibfield  {author} {\bibinfo {author} {\bibfnamefont {G.~H.}\ \bibnamefont
  {Lander}}, \bibinfo {author} {\bibfnamefont {M.~S.~S.}\ \bibnamefont
  {Brooks}}, \ and\ \bibinfo {author} {\bibfnamefont {B.}~\bibnamefont
  {Johansson}},\ }\href {https://link.aps.org/doi/10.1103/PhysRevB.43.13672}
  {\bibfield  {journal} {\bibinfo  {journal} {Physical Review B}\ }\textbf
  {\bibinfo {volume} {43}},\ \bibinfo {pages} {13672} (\bibinfo {year}
  {1991})}\BibitemShut {NoStop}%
\bibitem [{\citenamefont {Kaneko}\ \emph {et~al.}(2003)\citenamefont {Kaneko},
  \citenamefont {Metoki}, \citenamefont {Bernhoeft}, \citenamefont {Lander},
  \citenamefont {Ishii}, \citenamefont {Ikeda}, \citenamefont {Tokiwa},
  \citenamefont {Haga},\ and\ \citenamefont {Onuki}}]{Kaneko2003}%
  \BibitemOpen
  \bibfield  {author} {\bibinfo {author} {\bibfnamefont {K.}~\bibnamefont
  {Kaneko}}, \bibinfo {author} {\bibfnamefont {N.}~\bibnamefont {Metoki}},
  \bibinfo {author} {\bibfnamefont {N.}~\bibnamefont {Bernhoeft}}, \bibinfo
  {author} {\bibfnamefont {G.~H.}\ \bibnamefont {Lander}}, \bibinfo {author}
  {\bibfnamefont {Y.}~\bibnamefont {Ishii}}, \bibinfo {author} {\bibfnamefont
  {S.}~\bibnamefont {Ikeda}}, \bibinfo {author} {\bibfnamefont
  {Y.}~\bibnamefont {Tokiwa}}, \bibinfo {author} {\bibfnamefont
  {Y.}~\bibnamefont {Haga}}, \ and\ \bibinfo {author} {\bibfnamefont
  {Y.}~\bibnamefont {Onuki}},\ }\href
  {https://link.aps.org/doi/10.1103/PhysRevB.68.214419} {\bibfield  {journal}
  {\bibinfo  {journal} {Physical Review B}\ }\textbf {\bibinfo {volume} {68}},\
  \bibinfo {pages} {214419} (\bibinfo {year} {2003})}\BibitemShut {NoStop}%
\bibitem [{\citenamefont {Hiess}\ \emph {et~al.}(2001)\citenamefont {Hiess},
  \citenamefont {Boudarot}, \citenamefont {Coad}, \citenamefont {Brown},
  \citenamefont {Burlet}, \citenamefont {Lander}, \citenamefont {Brooks},
  \citenamefont {Kaczorowski}, \citenamefont {Czopnik},\ and\ \citenamefont
  {Troc}}]{Hiess2001}%
  \BibitemOpen
  \bibfield  {author} {\bibinfo {author} {\bibfnamefont {A.}~\bibnamefont
  {Hiess}}, \bibinfo {author} {\bibfnamefont {F.}~\bibnamefont {Boudarot}},
  \bibinfo {author} {\bibfnamefont {S.}~\bibnamefont {Coad}}, \bibinfo {author}
  {\bibfnamefont {P.~J.}\ \bibnamefont {Brown}}, \bibinfo {author}
  {\bibfnamefont {P.}~\bibnamefont {Burlet}}, \bibinfo {author} {\bibfnamefont
  {G.~H.}\ \bibnamefont {Lander}}, \bibinfo {author} {\bibfnamefont {M.~S.~S.}\
  \bibnamefont {Brooks}}, \bibinfo {author} {\bibfnamefont {D.}~\bibnamefont
  {Kaczorowski}}, \bibinfo {author} {\bibfnamefont {A.}~\bibnamefont
  {Czopnik}}, \ and\ \bibinfo {author} {\bibfnamefont {R.}~\bibnamefont
  {Troc}},\ }\href {http://stacks.iop.org/0295-5075/55/i=2/a=267
  http://iopscience.iop.org/article/10.1209/epl/i2001-00409-9/pdf} {\bibfield
  {journal} {\bibinfo  {journal} {EPL (Europhysics Letters)}\ }\textbf
  {\bibinfo {volume} {55}},\ \bibinfo {pages} {267} (\bibinfo {year}
  {2001})}\BibitemShut {NoStop}%
\bibitem [{\citenamefont {Maglic}\ \emph {et~al.}(1978)\citenamefont {Maglic},
  \citenamefont {Lander}, \citenamefont {Mueller},\ and\ \citenamefont
  {Kleb}}]{Maglic1978}%
  \BibitemOpen
  \bibfield  {author} {\bibinfo {author} {\bibfnamefont {R.~C.}\ \bibnamefont
  {Maglic}}, \bibinfo {author} {\bibfnamefont {G.~H.}\ \bibnamefont {Lander}},
  \bibinfo {author} {\bibfnamefont {M.~H.}\ \bibnamefont {Mueller}}, \ and\
  \bibinfo {author} {\bibfnamefont {R.}~\bibnamefont {Kleb}},\ }\href
  {https://link.aps.org/doi/10.1103/PhysRevB.17.308} {\bibfield  {journal}
  {\bibinfo  {journal} {Physical Review B}\ }\textbf {\bibinfo {volume} {17}},\
  \bibinfo {pages} {308} (\bibinfo {year} {1978})}\BibitemShut {NoStop}%
\bibitem [{\citenamefont {Hjelm}\ \emph {et~al.}(1993)\citenamefont {Hjelm},
  \citenamefont {Eriksson},\ and\ \citenamefont {Johansson}}]{Hjelm1993}%
  \BibitemOpen
  \bibfield  {author} {\bibinfo {author} {\bibfnamefont {A.}~\bibnamefont
  {Hjelm}}, \bibinfo {author} {\bibfnamefont {O.}~\bibnamefont {Eriksson}}, \
  and\ \bibinfo {author} {\bibfnamefont {B.}~\bibnamefont {Johansson}},\ }\href
  {https://link.aps.org/doi/10.1103/PhysRevLett.71.1459} {\bibfield  {journal}
  {\bibinfo  {journal} {Physical Review Letters}\ }\textbf {\bibinfo {volume}
  {71}},\ \bibinfo {pages} {1459} (\bibinfo {year} {1993})}\BibitemShut
  {NoStop}%
\bibitem [{\citenamefont {Hjelm}\ \emph {et~al.}(1994)\citenamefont {Hjelm},
  \citenamefont {Trygg}, \citenamefont {Eriksson}, \citenamefont {Johansson},\
  and\ \citenamefont {Wills}}]{Hjelm1994}%
  \BibitemOpen
  \bibfield  {author} {\bibinfo {author} {\bibfnamefont {A.}~\bibnamefont
  {Hjelm}}, \bibinfo {author} {\bibfnamefont {J.}~\bibnamefont {Trygg}},
  \bibinfo {author} {\bibfnamefont {O.}~\bibnamefont {Eriksson}}, \bibinfo
  {author} {\bibfnamefont {B.}~\bibnamefont {Johansson}}, \ and\ \bibinfo
  {author} {\bibfnamefont {J.}~\bibnamefont {Wills}},\ }\href
  {https://link.aps.org/doi/10.1103/PhysRevB.50.4332} {\bibfield  {journal}
  {\bibinfo  {journal} {Physical Review B}\ }\textbf {\bibinfo {volume} {50}},\
  \bibinfo {pages} {4332} (\bibinfo {year} {1994})}\BibitemShut {NoStop}%
\bibitem [{\citenamefont {Paix\~{a}o}\ \emph {et~al.}(1993)\citenamefont
  {Paix\~{a}o}, \citenamefont {Lander}, \citenamefont {Delapalme},
  \citenamefont {Nakotte}, \citenamefont {Boer},\ and\ \citenamefont
  {Br\"{u}ck}}]{Paixao1993}%
  \BibitemOpen
  \bibfield  {author} {\bibinfo {author} {\bibfnamefont {J.~A.}\ \bibnamefont
  {Paix\~{a}o}}, \bibinfo {author} {\bibfnamefont {G.~H.}\ \bibnamefont
  {Lander}}, \bibinfo {author} {\bibfnamefont {A.}~\bibnamefont {Delapalme}},
  \bibinfo {author} {\bibfnamefont {H.}~\bibnamefont {Nakotte}}, \bibinfo
  {author} {\bibfnamefont {F.~R.~d.}\ \bibnamefont {Boer}}, \ and\ \bibinfo
  {author} {\bibfnamefont {E.}~\bibnamefont {Br\"{u}ck}},\ }\href
  {http://stacks.iop.org/0295-5075/24/i=7/a=017
  http://iopscience.iop.org/article/10.1209/0295-5075/24/7/017/pdf} {\bibfield
  {journal} {\bibinfo  {journal} {EPL (Europhysics Letters)}\ }\textbf
  {\bibinfo {volume} {24}},\ \bibinfo {pages} {607} (\bibinfo {year}
  {1993})}\BibitemShut {NoStop}%
\bibitem [{\citenamefont {Brooks}\ \emph {et~al.}(1989)\citenamefont {Brooks},
  \citenamefont {Eriksson},\ and\ \citenamefont {Johansson}}]{Brooks1989}%
  \BibitemOpen
  \bibfield  {author} {\bibinfo {author} {\bibfnamefont {M.~S.~S.}\
  \bibnamefont {Brooks}}, \bibinfo {author} {\bibfnamefont {O.}~\bibnamefont
  {Eriksson}}, \ and\ \bibinfo {author} {\bibfnamefont {B.}~\bibnamefont
  {Johansson}},\ }\href {http://stacks.iop.org/0953-8984/1/i=34/a=004}
  {\bibfield  {journal} {\bibinfo  {journal} {Journal of Physics: Condensed
  Matter}\ }\textbf {\bibinfo {volume} {1}},\ \bibinfo {pages} {5861} (\bibinfo
  {year} {1989})}\BibitemShut {NoStop}%
\bibitem [{\citenamefont {Taupin}\ \emph {et~al.}(2015)\citenamefont {Taupin},
  \citenamefont {Sanchez}, \citenamefont {Brison}, \citenamefont {Aoki},
  \citenamefont {Lapertot}, \citenamefont {Wilhelm},\ and\ \citenamefont
  {Rogalev}}]{Taupin2015}%
  \BibitemOpen
  \bibfield  {author} {\bibinfo {author} {\bibfnamefont {M.}~\bibnamefont
  {Taupin}}, \bibinfo {author} {\bibfnamefont {J.~P.}\ \bibnamefont {Sanchez}},
  \bibinfo {author} {\bibfnamefont {J.~P.}\ \bibnamefont {Brison}}, \bibinfo
  {author} {\bibfnamefont {D.}~\bibnamefont {Aoki}}, \bibinfo {author}
  {\bibfnamefont {G.}~\bibnamefont {Lapertot}}, \bibinfo {author}
  {\bibfnamefont {F.}~\bibnamefont {Wilhelm}}, \ and\ \bibinfo {author}
  {\bibfnamefont {A.}~\bibnamefont {Rogalev}},\ }\href
  {https://link.aps.org/doi/10.1103/PhysRevB.92.035124} {\bibfield  {journal}
  {\bibinfo  {journal} {Physical Review B}\ }\textbf {\bibinfo {volume} {92}},\
  \bibinfo {pages} {035124} (\bibinfo {year} {2015})}\BibitemShut {NoStop}%
\bibitem [{\citenamefont {Proke\v{s}}\ \emph {et~al.}(2010)\citenamefont
  {Proke\v{s}}, \citenamefont {de~Visser}, \citenamefont {Huang}, \citenamefont
  {F\r{a}k},\ and\ \citenamefont {Ressouche}}]{Prokes2010}%
  \BibitemOpen
  \bibfield  {author} {\bibinfo {author} {\bibfnamefont {K.}~\bibnamefont
  {Proke\v{s}}}, \bibinfo {author} {\bibfnamefont {A.}~\bibnamefont
  {de~Visser}}, \bibinfo {author} {\bibfnamefont {Y.~K.}\ \bibnamefont
  {Huang}}, \bibinfo {author} {\bibfnamefont {B.}~\bibnamefont {F\r{a}k}}, \
  and\ \bibinfo {author} {\bibfnamefont {E.}~\bibnamefont {Ressouche}},\ }\href
  {https://link.aps.org/doi/10.1103/PhysRevB.81.180407} {\bibfield  {journal}
  {\bibinfo  {journal} {Physical Review B}\ }\textbf {\bibinfo {volume} {81}},\
  \bibinfo {pages} {180407} (\bibinfo {year} {2010})}\BibitemShut {NoStop}%
\bibitem [{\citenamefont {Kernavanois}\ \emph {et~al.}(2001)\citenamefont
  {Kernavanois}, \citenamefont {Grenier}, \citenamefont {Huxley}, \citenamefont
  {Ressouche}, \citenamefont {Sanchez},\ and\ \citenamefont
  {Flouquet}}]{Kernavanois2001}%
  \BibitemOpen
  \bibfield  {author} {\bibinfo {author} {\bibfnamefont {N.}~\bibnamefont
  {Kernavanois}}, \bibinfo {author} {\bibfnamefont {B.}~\bibnamefont
  {Grenier}}, \bibinfo {author} {\bibfnamefont {A.}~\bibnamefont {Huxley}},
  \bibinfo {author} {\bibfnamefont {E.}~\bibnamefont {Ressouche}}, \bibinfo
  {author} {\bibfnamefont {J.~P.}\ \bibnamefont {Sanchez}}, \ and\ \bibinfo
  {author} {\bibfnamefont {J.}~\bibnamefont {Flouquet}},\ }\href
  {https://link.aps.org/doi/10.1103/PhysRevB.64.174509} {\bibfield  {journal}
  {\bibinfo  {journal} {Physical Review B}\ }\textbf {\bibinfo {volume} {64}},\
  \bibinfo {pages} {174509} (\bibinfo {year} {2001})}\BibitemShut {NoStop}%
\bibitem [{\citenamefont {Burlet}\ \emph {et~al.}(1994)\citenamefont {Burlet},
  \citenamefont {Troc}, \citenamefont {Kaczorowski}, \citenamefont {Noël},\
  and\ \citenamefont {Rossat-Mignod}}]{Burlet1994}%
  \BibitemOpen
  \bibfield  {author} {\bibinfo {author} {\bibfnamefont {P.}~\bibnamefont
  {Burlet}}, \bibinfo {author} {\bibfnamefont {R.}~\bibnamefont {Troc}},
  \bibinfo {author} {\bibfnamefont {D.}~\bibnamefont {Kaczorowski}}, \bibinfo
  {author} {\bibfnamefont {H.}~\bibnamefont {Noël}}, \ and\ \bibinfo {author}
  {\bibfnamefont {J.}~\bibnamefont {Rossat-Mignod}},\ }\href {\doibase
  https://doi.org/10.1016/0304-8853(94)90678-5} {\bibfield  {journal} {\bibinfo
   {journal} {Journal of Magnetism and Magnetic Materials}\ }\textbf {\bibinfo
  {volume} {130}},\ \bibinfo {pages} {237} (\bibinfo {year}
  {1994})}\BibitemShut {NoStop}%
\end{thebibliography}%


\begin{thebibliography}{0}%
\makeatletter
\providecommand \@ifxundefined [1]{%
 \@ifx{#1\undefined}
}%
\providecommand \@ifnum [1]{%
 \ifnum #1\expandafter \@firstoftwo
 \else \expandafter \@secondoftwo
 \fi
}%
\providecommand \@ifx [1]{%
 \ifx #1\expandafter \@firstoftwo
 \else \expandafter \@secondoftwo
 \fi
}%
\providecommand \natexlab [1]{#1}%
\providecommand \enquote  [1]{``#1''}%
\providecommand \bibnamefont  [1]{#1}%
\providecommand \bibfnamefont [1]{#1}%
\providecommand \citenamefont [1]{#1}%
\providecommand \href@noop [0]{\@secondoftwo}%
\providecommand \href [0]{\begingroup \@sanitize@url \@href}%
\providecommand \@href[1]{\@@startlink{#1}\@@href}%
\providecommand \@@href[1]{\endgroup#1\@@endlink}%
\providecommand \@sanitize@url [0]{\catcode `\\12\catcode `\$12\catcode
  `\&12\catcode `\#12\catcode `\^12\catcode `\_12\catcode `\%12\relax}%
\providecommand \@@startlink[1]{}%
\providecommand \@@endlink[0]{}%
\providecommand \url  [0]{\begingroup\@sanitize@url \@url }%
\providecommand \@url [1]{\endgroup\@href {#1}{\urlprefix }}%
\providecommand \urlprefix  [0]{URL }%
\providecommand \Eprint [0]{\href }%
\providecommand \doibase [0]{http://dx.doi.org/}%
\providecommand \selectlanguage [0]{\@gobble}%
\providecommand \bibinfo  [0]{\@secondoftwo}%
\providecommand \bibfield  [0]{\@secondoftwo}%
\providecommand \translation [1]{[#1]}%
\providecommand \BibitemOpen [0]{}%
\providecommand \bibitemStop [0]{}%
\providecommand \bibitemNoStop [0]{.\EOS\space}%
\providecommand \EOS [0]{\spacefactor3000\relax}%
\providecommand \BibitemShut  [1]{\csname bibitem#1\endcsname}%
\let\auto@bib@innerbib\@empty
\end{thebibliography}%

\end{document}